\documentclass[11pt,onecolumn]{article}
\usepackage{amsmath,amsthm,amssymb,amsfonts,latexsym,epsfig,graphicx}
\usepackage{algorithm,algorithmicx}
\usepackage{color}
\usepackage{dsfont}
\usepackage{wrapfig}

\usepackage[left=0.9in,top=1.0in,right=0.9in,bottom=1.0in,nohead,textheight=10in,footskip=0.3in]{geometry}

\newtheorem{theorem}{Theorem}
\newtheorem{corollary}[theorem]{Corollary}
\newtheorem{conjecture}{Conjecture}
\newtheorem{lemma}[theorem]{Lemma}
\newtheorem{definition}[theorem]{Definition}
\newtheorem{proposition}[theorem]{Proposition}
\newtheorem{remark}{Remark}
\newtheorem{example}{Example}

\DeclareMathOperator{\supp}{supp}
\DeclareMathOperator{\dom}{dom}

\DeclareMathOperator{\ar}{ar}
\DeclareMathOperator{\Pol}{Pol}

\newcommand{\q}{\ensuremath{\mathbb{Q}}}

\newcommand{\fPol}[1]{\ensuremath{\operatorname{fPol}(#1)}}
\newcommand{\Polplus}[1]{\ensuremath{\operatorname{Pol}^+(#1)}}
\DeclareMathOperator{\pol}{Pol}
\DeclareMathOperator{\mwis}{MWIS}

\DeclareMathOperator{\CostF}{{\bf \Phi}}

\newcommand{\CSP}[1]{\ensuremath{\operatorname{CSP}(#1)}}
\newcommand{\cspo}{\textsc{CSP}}
\newcommand{\vcspo}{\textsc{VCSP}}
\newcommand{\vcsp}[1]{\ensuremath{\operatorname{VCSP}(#1)}}
\newcommand{\VCSP}[1]{\ensuremath{\operatorname{VCSP}(#1)}}

\begin{document}

\def\myparagraph#1{\vspace{2pt}\noindent{\bf #1~~}}


\newif\ifTR\TRtrue

\long\def\ignore#1{}
\def\myps[#1]#2{\includegraphics[#1]{#2}}
\def\etal{{\em et al.}}
\def\Bar#1{{\bar #1}}
\def\br(#1,#2){{\langle #1,#2 \rangle}}
\def\setZ[#1,#2]{{[ #1 .. #2 ]}}
\def\Pr{\mbox{\rm Pr}}
\def\zd{,\ldots,}

\newcommand{\Z}{\mbox{$\mathbb Z$}}
\newcommand{\R}{\mbox{$\mathbb R$}}
\newcommand{\Q}{\mbox{$\mathbb Q$}}
\newcommand{\Qnn}{\mbox{$\mathbb Q_{\geq 0}$}}
\newcommand{\Qnnc}{\mbox{$\overline{\mathbb Q}_{\geq 0}$}}
\newcommand{\Qc}{\mbox{$\overline{\mathbb Q}$}}
\newcommand{\Rnn}{\mbox{$\mathbb R_{\geq 0}$}}
\newcommand{\Rnnc}{\mbox{$\overline{\mathbb R}_{\geq 0}$}}
\newcommand{\Rc}{\mbox{$\overline{\mathbb R}$}}
\newcommand{\N}{\mbox{$\mathbb N$}}

\def\closure#1{{\langle#1\rangle}}
\def\setof#1{{\left\{#1\right\}}}
\def\suchthat#1#2{\setof{\,#1\mid#2\,}} 
\def\event#1{\setof{#1}}
\def\q={\quad=\quad}
\def\qq={\qquad=\qquad}
\def\calA{{\cal A}}
\def\calC{{\cal C}}
\def\calD{{\cal D}}
\def\calE{{\cal E}}
\def\calF{{\cal F}}
\def\calG{{\cal G}}
\def\calI{{\cal I}}
\def\calH{{\cal H}}
\def\calL{{\cal L}}
\def\calN{{\cal N}}
\def\calP{{\cal P}}
\def\calR{{\cal R}}
\def\calS{{\cal S}}
\def\calT{{\cal T}}
\def\calU{{\cal U}}
\def\calV{{\cal V}}
\def\calO{{\cal O}}
\def\calX{{\cal X}}
\def\s{\footnotesize}
\def\calNG{{\cal N_G}}
\def\psfile[#1]#2{}
\def\psfilehere[#1]#2{}
\def\epsfw#1#2{\includegraphics[width=#1\hsize]{#2}}
\def\assign(#1,#2){\langle#1,#2\rangle}
\def\edge(#1,#2){(#1,#2)}
\def\VS{\calV^s}
\def\VT{\calV^t}
\def\slack(#1){\texttt{slack}({#1})}
\def\barslack(#1){\overline{\texttt{slack}}({#1})}
\def\NULL{\texttt{NULL}}
\def\PARENT{\texttt{PARENT}}
\def\GRANDPARENT{\texttt{GRANDPARENT}}
\def\TAIL{\texttt{TAIL}}
\def\HEADORIG{\texttt{HEAD$\_\:$ORIG}}
\def\TAILORIG{\texttt{TAIL$\_\:$ORIG}}
\def\HEAD{\texttt{HEAD}}
\def\CURRENTEDGE{\texttt{CURRENT$\!\_\:$EDGE}}

\def\unitvec(#1){{{\bf u}_{#1}}}
\def\uvec{{\bf u}}
\def\vvec{{\bf v}}
\def\Nvec{{\bf N}}

\newcommand{\bg}{\mbox{$\bf g$}}
\newcommand{\bh}{\mbox{$\bf h$}}

\newcommand{\bw}{\mbox{\boldmath $w$}}
\newcommand{\bvarphi}{\mbox{\boldmath $\varphi$}}

\newcommand\myqed{{}}

\newcommand{\bGamma}{\mathbf{\Gamma}}
\newcommand{\bR}{\mathbf{R}}
\newcommand{\bv}{\mathbf{v}}
\newcommand{\bu}{\mathbf{u}}
\newcommand{\bx}{\mathbf{x}}
\newcommand{\by}{\mathbf{y}}



\title{\Large\bf  
Effectiveness of Structural Restrictions for Hybrid CSPs}

\author{
     Vladimir Kolmogorov$^{\tiny \dag}$ \\ {\normalsize\tt vnk@ist.ac.at}
\and Michal Rol\'inek$^{\tiny \dag}$ \\ {\normalsize\tt michal.rolinek@ist.ac.at}
\and Rustem Takhanov$^{\tiny \ddag}$ \\ {\normalsize\tt takhanov@mail.ru}
\and \\ \\
\normalsize ~$^{\tiny \dag}$Institute of Science and Technology Austria \qquad
\normalsize ~$^{\tiny \ddag}$Nazarbayev University, Kazakhstan
}

\date{}
\maketitle


\begin{abstract}
Constraint Satisfaction Problem (CSP) is a fundamental algorithmic problem that appears
in many areas of Computer Science. It can be equivalently stated as computing a homomorphism
$\mbox{$\bR \rightarrow \bGamma$}$ between two relational structures, e.g.\ between two directed graphs.
Analyzing its complexity has been a prominent research direction, especially for the {\em fixed template CSPs}
where the right side $\bGamma$ is fixed and the left side $\bR$ is unconstrained. 

Far fewer results are known for the {\em hybrid} setting that restricts both sides simultaneously.
It assumes that $\bR$ belongs to a certain class of relational structures (called a {\em structural restriction} in this paper).
We study which structural restrictions are {\em effective}, i.e.\ there exists a fixed template $\bGamma$
(from a certain class of languages) for which the problem is tractable when $\bR$ is restricted, and NP-hard otherwise.
We provide a characterization for structural restrictions that are {\em closed under inverse homomorphisms}.
The criterion is based on the {\em chromatic number} of a relational structure defined in this paper; it generalizes the standard
chromatic number of a graph.

As our main tool, we use the algebraic machinery developed for fixed template CSPs.
To apply it to our case, we introduce a new construction called a ``lifted language''.
We also give a characterization for structural restrictions corresponding to minor-closed families of graphs,
extend results to certain Valued CSPs (namely conservative valued languages),
and state implications
for (valued) CSPs with ordered variables and
for the maximum weight independent set problem on some restricted families of graphs.
\end{abstract}


\section{Introduction}\label{sec:intro}

The {\it Constraint satisfaction problems (CSPs)} and the valued constraint satisfaction problems (VCSP) provide a powerful framework for analysis of a large set of computational problems arising in propositional logic, combinatorial optimization, artificial intelligence,  graph theory, scheduling, biology, computer vision etc. Traditionally CSP is formalized either as a problem of (a) finding an assignment of values to a given set of variables, subject to constraints on the values that can be assigned simultaneously to specified subsets of variables, or as problem of (b) finding a homomorphism between two finite relational structures $A$ and $B$ (e.g., two oriented graphs). 
These two formulations are polynomially equivalent under the condition that the input constraints in the first case or input relations in the second case are given by lists of their elements. Soft version of CSP, that is VCSP, generalizes the CSP by replacing crisp constraints with cost functions applied to tuples of variables. In the VCSP we require to find the maximum (or minimum) of a sum of cost functions applied to corresponding variables.

The CSPs have been the cutting edge research field of theoretical computer science since the 70s, and
recently this interest has been expanded to VCSP. One of the themes that revealed 
rich logical and algebraic structure of the CSPs was the question of classification of the problem's computational complexity when constraint relations are restricted to a given set of relations or, alternatively, when the second relational structure is some fixed $\bGamma$. Thus, this problem is parameterized by $\bGamma$, denoted as $\CSP\bGamma$ and called a fixed template CSP with a template $\bGamma$ (another name is a non-uniform CSP).
E.g., if the domain set is boolean and $\bGamma$ is a relational structure with four ternary predicates $x\vee y\vee z$,
$\overline{x}\vee y\vee z$, $\overline{x}\vee \overline{y}\vee z$,
$\overline{x}\vee \overline{y}\vee \overline{z}$, $\CSP\bGamma$ models 3-SAT which is historically one the first NP-complete problems~\cite{Cook:1971}. At the same time, if we restrict $\bGamma$ to binary predicates, then we obtain tractable 2-SAT. Generally, Schaeffer proved~\cite{schaefer78:complexity} that for any template $\bGamma$ over the boolean set,  $\CSP\bGamma$ is either in P or NP-complete, and any tractable constraint language belongs to one of 6 classes (0 or 1-preserving, binary, horn, anti-horn and linear subspaces). When $\bGamma$ contains only one graph (irreflexive symmetric predicate) Hell and Ne\v{s}et\v{r}il~\cite{Hell} proved an analogous statement, by showing that only for bipartite graphs the problem is tractable. Feder and Vardi~\cite{feder98:monotone} found that all fixed template CSPs can be expressed as problems in a fragment of SNP, called Monotone Monadic SNP (MM SNP). They introduced this class as a natural restriction of SNP for which Ladner's argument about the existence of problems with intermediate complexity between P and NP-hard could not be applied. Moreover, they showed that all problems in MM SNP can be reduced with respect to Turing reduction to fixed template CSPs and, thus, non-uniform CSPs complexity classification would lead to a classification of MM SNP problems. This result placed fixed-template CSPs into a broad logical context that naturally lead to a conjecture that such CSPs are either tractable or NP-hard, the so called dichotomy conjecture.

In \cite{Jeavons:1998} Jeavons observed that any predicate given by primitive positive formula using predicates of the template $\bGamma$, when added to $\bGamma$, does not change the complexity of $\CSP\bGamma$.
This result clarified that the computational complexity of $\CSP\bGamma$ is 
fully defined by the minimal predicate clone that contains predicates of $\bGamma$. In universal algebra, it has long been known that the predicate clones are dual to the so called functional clones \cite{Post41,kuznetsov,geiger1968}. Specifically, it implies that the complexity of $\CSP\bGamma$ is defined by the set of polymorphisms of $\bGamma$. The last was the main motive for subsequent research. Intensive studies in this direction lead to a conjectured algebraic description of all tractable templates made by Bulatov, Jeavons, and Krokhin \cite{bulatov05:classifying}, with subsequent reformulations of this conjecture by Maroti and McKenzie \cite{MarotiMcKenzie}. In the long run it was shown by Siggers~\cite{Siggers} that if Bulatov-Jeavons-Krokhin characterization of tractable templates is correct, then the tractable core structures can be characterized as those that admit a single 6-ary polymorphism that satisfies a certain equality. The last fact will serve as a key ingredient for one of our results.

Besides fixed template CSPs, another parameterization of CSP concerns restrictions on the left relational structure of the input. If we restrict the left structure of the input to some specified set $\mathcal{H}$ and impose no restriction on the right relational structure, then the problem is called CSP with structural restrictions $\mathcal{H}$.
For example, if $\mathcal{H}$ is a set of graphs with treewidth less or equal to $k\in \mathbb{N}$, then the problem can be solved in polynomial time. It was found by Grohe~\cite{Grohe:2007} that any structural restriction $\mathcal{H}$ that defines tractable CSP should be of bounded treewidth modulo homomorphic equivalence.

{\bf Related work.} Since many (V)CSP instances do not fall into any of the tractable classes offered by one of the previous approaches, there has been growing interest in the so-called hybrid restrictions. That is when the input is restricted to a subset of all input pairs $\left(\bR,\bGamma\right)$. One approach to this problem is to construct a new structure for any input $\left(\bR,\bGamma\right)$, $G_{\bR,\bGamma}$, and shift the analysis to $G_{\bR,\bGamma}$. In case of binary CSPs (i.e.\ when all predicates of an input are binary) it is natural to define $G_{\bR,\bGamma}$ as a microstructure graph \cite{Jegou93a} of a template $\left(\bR,\bGamma\right)$. Thereby, a set of inputs for which certain local substructures in $G_{\bR,\bGamma}$ are forbidden form a parametrized problem. Cooper and {\v{Z}}ivn\'y \cite{cz11:ai} investigated this formulation and found examples of specific forbidden substructures that result in tractable hybrid CSPs. Microstructure graphs also naturally appear in the context of fixed template CSPs. Specifically, all templates $\bGamma$ with binary predicates that define fixed template CSPs for which local consistency preprocessing of the input results in a perfect microstructure graph were completely classified in \cite{Takhanov10adichotomy}.

{\bf Our results.} The main topic of our paper is a hybrid framework for (V)CSP, when left structures are restricted to some set $\mathcal{H}$ and combined with a fixed right structure $\bGamma$ (corresponding CSP is denoted as $\cspo_{\mathcal{H}}( \bGamma )$). The difficulty of applying known algebraic machinery to this framework is due to the fact that the closure operator, analogous to the minimal containing clone, cannot depend on $\bGamma$ only. Therefore, in an algebraic theory of hybrid CSPs an analogue of primitive positive formula should depend on both input structures. 
In our approach we define for any $\bR\in \mathcal{H}$ and $\bGamma$ a set of predicates $\bGamma_{\bR}$ that we call  
\ifTR
a ``lifted'' language (see Sec.~\ref{sec:construction}). 
\else
a ``lifted'' language.
\fi
Our key idea is that the closures $\langle\bGamma_{\bR}\rangle$ for $\bR\in \mathcal{H}$, under certain conditions, could maintain the information on the tractability of $\cspo_{\mathcal{H}}( \bGamma )$. In this paper, by that ``certain conditions'' we understand the property that $ \mathcal{H}$ is closed under inverse homomorphisms. We are especially interested in a classification of structural restrictions $ \mathcal{H}$ closed under inverse homomorphisms for which we could find a template $\bGamma$ (in a certain class of templates $\mathcal{C}$) that defines tractable 
$\cspo_{\mathcal{H}}( \bGamma )$, whereas a $\CSP\bGamma$ is NP-hard. We call such restrictions effective for a class $\mathcal{C}$. Our key results are formulated for 2 cases: the class of BJK languages, that is, the class of templates that are either tractable or have core a without a Siggers polymorphism, and a class of conservative valued templates.

Specifically, we prove that if $ \mathcal{H}$ is a set of binary structures closed under inverse homomorphisms, it is effective for BJK languages if and only if $\left\{\chi(\bR)\:|\:\bR\in \mathcal{H}\right\}$ is bounded, where $\chi(\bR)$ is a chromatic number of $\bR$ (considered as a graph). The last result is extended to the nonbinary case, with natural generalization of the chromatic number to arbitrary relational structures.
A notable corollary of this result is that the set of acyclic digraphs is an ineffective structural restriction for BJK languages. This explains why NP-hardness arguments for certain fixed templates of digraph homomorphism problem can be extended to a case when the input digraph is acyclic \cite{swarts}. Less straightforward corollary: let $\mathcal{H}$ be a set of binary structures such that their ``graph copies'' forbid specific minors, then $\mathcal{H}$ is effective for BJK languages if and only if $\left\{\chi(\bR)\:|\:\bR\in \mathcal{H}\right\}$ is bounded. The last statement does not require that $\mathcal{H}$ is closed under inverse homomorphisms. 

For $\vcspo_{\mathcal{H}}( \bGamma )$ we prove an analogue of our previous result for a class $\mathcal{C}$ of all conservative valued templates. We obtain as a corollary that the maximum weight independent set problem is still NP-hard in some graph classes. 


\ifTR

{\bf Organization.} In Sec.~\ref{sec:prelim} we give all the preliminary definitions and state theorems that we need.
In Sec.~\ref{sec:result} we state our main results (Theorems \ref{coloreffective}, \ref{bjkineffective} and \ref{vcspconsineff})
and their implications. The proofs of the main theorems are given in Sec.~\ref{sec:effproof}-\ref{sec:vcspcons}.

\fi

\section{Preliminaries}\label{sec:prelim}

Throughout the paper we assume $P \neq NP$. A problem is called {\em tractable} if it can be solved in polynomial time.

%
%
%

The symbol $[n]$ will denote the set $\{1, \dots, n\}$, and $\Qc=\mathbb{Q}\cup\{\infty\}$ the set
of rational numbers with (positive) infinity. Also $D$ will stand for a finite set.

We will denote the tuples in lowercase boldface such as $\mathbf{a} = (a_1, \dots, a_k)$. Also for mappings $h \colon A \to B$ and tuples $\mathbf{a} = (a_1, \dots, a_k)$, where $a_j \in A$ for $j = 1, \dots, k$, we will write $\mathbf{b} = (h(a_1), \dots, h(a_k))$ simply as $\mathbf{b} = h(\mathbf{a})$. Relational structures will be denoted in uppercase boldface as $\mathbf{R} = (R, r_1, \dots, r_k)$.

Finally let $\ar(\varrho)$, $\ar(\mathbf{a})$, and $\ar(f)$ stand for arity of a relation $\varrho$, size of a tuple $\mathbf{a}$, and arity (number of parameters) of a function $f$, respectively.

\subsection{Fixed template CSP}

We will first formulate the general CSP in an algebraic way as a decision problems whether there exists a homomorphism between certain relational structures.

\begin{definition} Let $\mathbf{R} = (R, r_1, \dots, r_k)$ and $\mathbf{R}' = (R', r'_1, \dots, r'_k)$ be relational structures with a common signature (that is $\ar(r_i)= \ar(r'_i)$  for every $i = 1, \dots, k$). A mapping $h\colon R \to R'$ is called a \emph{homomorphism} from $\mathbf{R}$ to $\mathbf{R}'$ if for each $i = 1, \dots, k$, whenever $(x_1, \dots, x_{\ar(r_i)}) \in r_i$, then $((h(x_1), \dots, h(x_{\ar(r'_i)})) \in r'_i$. In that case, we write $\mathbf{R} \stackrel{h}{\to} \mathbf{R}'$ or sometimes just $\mathbf{R} \to \mathbf{R}'$.
\end{definition}

\begin{definition}[General CSP] The general CSP is the following decision problem. Given a pair of relational structures with common signature $\mathbf{R} = (V,r_1, \dots, r_k)$ and $\bGamma = (D, \varrho_1, \dots, \varrho_k)$, decide whether $\mathbf{R} \to \bGamma$.
Equivalently,  decide whether there is a mapping $h:V\rightarrow D$ that satisfies
\begin{equation}
\bigwedge_{(\varrho,\bv)\in T}[h(\bv)\in\varrho]
\end{equation}
where $T=\{(\varrho_i,\bv)\:|\:i\in[k],\bv\in r_i\}$ specifies the set of constraints.
\end{definition}

The set $V$ represents the set of \emph{variables} and we will only consider $V$ finite, similarly $D$ is the \emph{domain set} or the set of \emph{labels} for variables. The relations $r_1$, \dots, $r_k$ specify the tuples of $V$ constrained by relations $\varrho_1$, \dots, $\varrho_k$, respectively.

As we mentioned in the introduction, one natural way to restrict the general CSP is to fix the constraint types. A finitary relational structure $\mathbf{\Gamma} = (D, \varrho_1, \dots, \varrho_k)$ over a fixed finite domain $D$ will be called a constraint language.
For such $\bGamma$ we will denote by $\Gamma$ (without boldface) the set of relations $\{ \varrho_1, \dots, \varrho_k \}$; with some abuse of
terminology set $\Gamma$ will also be called a constraint language. (Note that both views are used in the literature).




\begin{definition}[Fixed template CSP] Let $D$ be a finite set and $\bGamma$ a constraint language over $D$. Then the decision problem $\CSP\bGamma$ is defined as follows:
given a relational structure $\mathbf{R} = (V,r_1, \dots, r_k)$ of the same signature as $\bGamma$, decide whether $\mathbf{R} \to \mathbf{\Gamma}$.
\end{definition}


We will usually write $\CSP\Gamma$ instead of $\CSP\bGamma$.
Although there are multiple relational structures $\bGamma$ that correspond to the same set $\Gamma$,
it can be seen that all choices give  equivalent problems; this justifies the notation  $\CSP\Gamma$.

\subsection{Fixed template VCSP}

A more general framework operates with {\em cost functions} $f:D^n \rightarrow \Qc$ instead of relations $\varrho\subseteq D^n$. This idea leads to the notion of valued CSP.

\begin{definition}
We denote the set of all functions $f:D^n \rightarrow \Qc$ by $\CostF_D^{(n)}$ and let $\CostF_D=\bigcup_{n\ge 1}{\CostF_D^{(n)}}$.
We will often call the functions in $\CostF_D$ {\em cost functions} over $D$.
For every cost function $f\in\CostF_D^{(n)}$, let $\dom f=\{x\mid f(x)<\infty\}$. Note that $\dom f$ can be considered both as an $n$-ary relation
and as an $n$-ary function such that $\dom f(x)=0$ if and only if $f(x)$ is finite.
\end{definition}

We will say that the cost functions in $\CostF_D$ take \emph{values}.
\ifTR
Note that in some papers on VCSP, e.g.~\cite{cccjz13:sicomp,Thapper15:Sherali}, cost functions are called weighted relations.
\else
Note that in some papers on VCSP cost functions are called weighted relations.
\fi

\begin{definition}
An instance of the {\em valued constraint satisfaction problem} (VCSP)
is specified by finite sets $D$, $V$ and a function from $D^V$ to $\Qc$ given by
\begin{equation}\label{eq:VCSPinst}
f_{\calI}(h)=\sum_{(f,\bv)\in T}{w(f,\bv)f(h(\bv))},
\end{equation}
where $V$ is a finite set of variables, $w(f,\bv)$ are positive numbers,\footnote{
We will allow two possibilities: (i) weights are positive integers, and the length of the description of $\calI$ grows linearly with $w(f,\bv)$;
(ii) weights are positive rationals. All our statements for VCSPs will hold under both models.
Note that in the literature weights $w(f,\bv)$ are usually omitted, and $T$ is allowed to be a multiset rather than a set;
this is equivalent to model (i). Including weights will be convenient for hybrid VCSPs. }
and $T$ is a finite set of constraints of the form $(f,\bv)$
where $f \in \CostF_D$ is a cost function and $\bv\in V^{\ar(f)}$ is a tuple of variables of size $\ar(f)$.
The goal is to find an {\em assignment} (or {\em labeling}) $h\in D^V$ that minimizes $f_\calI$.
\end{definition}

Note that $f_\calI$ can also be looked at as a cost function over the variable set $V$. 

\begin{definition}
A \emph{valued constraint language} over $D$ is either 
a tuple $\bGamma=(D,f_1,\ldots,f_k)$ with $f_1,\ldots,f_k\in \CostF_D$
or the corresponding finite set $\Gamma=\{f_1,\ldots,f_k\}\subseteq \CostF_D$.
We will denote by \vcsp{\Gamma} the class of all VCSP instances in which the
cost functions are all contained in $\Gamma$.
\end{definition}

This framework subsumes many other frameworks studied earlier and captures many specific well-known problems,
including {\sc $k$-Sat}, {\sc Graph $k$-Colouring}, {\sc Max Cut}, {\sc Min Vertex Cover}, and others (see~\cite{jeavons14:beatcs}).

A function $f\in\CostF_D^{(n)}$ that takes values in $\{0,\infty\}$ is called {\em crisp}. We will often view it as a relation in $D^n$,
and vice versa
(this should be clear from the context). If language $\Gamma$ is crisp (i.e.\ it contains only crisp functions),
then $\VCSP\Gamma$ is a pure feasibility problem corresponding to $\CSP\Gamma$. Note, however, that according to our definitions there is
a slight difference between the two: $\CSP\Gamma$ is a decision problem while $\VCSP\Gamma$ asks to compute a solution explicitly if it exists.


The dominant research line in this area is to classify the complexity of problems $\VCSP\Gamma$. 
Sometimes, problems $\CSP\Gamma$ and $\VCSP\Gamma$ are 
defined also for infinite languages $\Gamma$ and then $\VCSP\Gamma$ is called tractable if for each finite $\Gamma'\subseteq \Gamma$, $\VCSP{\Gamma'}$ is tractable. Also, $\VCSP\Gamma$ is called NP-hard if for some finite $\Gamma'\subseteq \Gamma$, $\VCSP{\Gamma'}$ is NP-hard. In turn, we will focus purely on finite languages $\Gamma$.

\ifTR
\subsection{Polymorphisms, Expressibility}
\else
\subsection{Polymorphisms}
\fi

Let $\calO_D^{(m)}$ denote the set of all operations $g:D^m\rightarrow D$ and let
 $\calO_D=\bigcup_{m\ge 1}{\calO_D^{(m)}}$. When $D$ is clear from the context, we will sometimes write simply $\calO^{(m)}$ and $\calO$.

Any language $\Gamma$ defined on $D$ can be associated
with a set of operations on $D$,
known as the polymorphisms of $\Gamma$, defined as follows.
\begin{definition}
\label{def:polymorphism}
An operation $g\in \calO_D^{(m)}$ is a \emph{polymorphism} of a cost function $f \in \CostF_D$ if
for any $\mathbf{x}^1,\ldots,\mathbf{x}^m \in \dom f$,
we have that $g(\mathbf{x}^1,\ldots,\mathbf{x}^m)\in \dom f$ where $g$ is applied component-wise.

For any valued constraint language $\Gamma$ over a set $D$,
we denote by $\pol(\Gamma)$ the set of all operations on $D$ that are polymorphisms of every
$f \in \Gamma$.
\end{definition}
%

Clearly, if $g$ is a polymorphism of a cost function $f$, then $g$ is also a polymorphism of $\dom f$.
For $\{0,\infty\}$-valued functions, which naturally correspond to relations, the notion
of a polymorphism defined above coincides with the standard notion of a polymorphism for relations.
Note that the projections, i.e. operations of the form $e_n^i(x_1,\ldots,x_n)=x_i$,  are polymorphisms of all valued constraint languages.
\ifTR 
Polymorphisms play the key role in the algebraic approach to the CSP, but for VCSPs more general constructs are necessary. Now we define them.

\begin{definition}
An $m$-ary \emph{fractional operation} $\omega$ on $D$ is a probability distribution on $\calO_D^{(m)}$.
The support of $\omega$ is defined as $\supp(\omega)=\{g\in \calO_D^{(m)}\mid \omega(g)>0\}$.
\end{definition}

\begin{definition}\label{def:fpol}
A $m$-ary fractional operation $\omega$ on $D$ is said to be a \emph{fractional polymorphism} of a
cost function $f\in \CostF_D$ if for any $\mathbf{x}^1,\ldots,\mathbf{x}^m \in \dom f$,
we have
\begin{equation}
\sum_{g\in\supp(\omega)}{\omega(g)f(g(\mathbf{x}^1,\ldots,\mathbf{x}^m))} \le \frac{1}{m}(f(\mathbf{x}^1)+\ldots+f(\mathbf{x}^m)).
\label{eq:wpol-dist}
\end{equation}

For a constraint language $\Gamma$, $\fPol\Gamma$ will denote the set of all fractional operations that are fractional polymorphisms
of each function in $\Gamma$. Also, let $\Polplus\Gamma=\{g\in \calO_D\mid g\in\supp(\omega), \omega\in\fPol\Gamma\}$.
\end{definition}

Clearly, we have $\Polplus\Gamma\subseteq\Pol\Gamma$ for any $\Gamma$.

The key observation in the algebraic approach to (V)CSP is that
neither the complexity nor the algebraic properties of a language $\Gamma$ change when functions ``expressible'' from $\Gamma$ in a certain way are added to it.

\begin{definition}
For a constraint language $\Gamma$, let $\langle\Gamma\rangle$ denote the set of all functions $f(x_1\zd x_k)$ such that,
for some instance $\calI$ of $\VCSP\Gamma$ with objective function $f_{\calI}(x_1\zd x_k,x_{k+1}\zd x_n)$,
we have \[f(x_1\zd x_k)=\min_{x_{k+1}\zd x_n}{f_{\calI}(x_1\zd x_k,x_{k+1}\zd x_n)}.\]
We then say that $\Gamma$ {\em expresses} $f$, and call $\langle\Gamma\rangle$ the {\em expressive power} of $\Gamma$.
\end{definition}

\begin{lemma}[\cite{Cohen06:algebraic,cohen06:complexitysoft}]
\label{lem:exppower}
Let $f\in \langle\Gamma\rangle$. Then
\begin{enumerate}
\item[(a)] if $\omega\in\fPol\Gamma$, then $\omega$ is a fractional polymorphism of $f$ and of $\dom f$;
\item[(b)] $\VCSP\Gamma$ is tractable if and only if $\VCSP{\Gamma\cup\{f,\dom f\}}$ is tractable;
\item[(c)] $\VCSP\Gamma$ is NP-hard if and only if $\VCSP{\Gamma\cup\{f,\dom f\}}$ is NP-hard.
\end{enumerate}
\end{lemma}
\else 
Polymorphisms play the key role in the algebraic approach to the CSP. For VCSPs more general constructs called {\em fractional polymorphisms} are necessary.
We refer to the full version of the paper~\cite{KRT:arxiv15} for further background on this topic.
\fi

\subsection{Algebraic dichotomy conjecture}
The condition for tractability of CSPs was first conjectured by Bulatov, Krokhin, and Jeavons~\cite{bulatov05:classifying},
and a number of equivalent formulations was later given in~\cite{Siggers,MarotiMcKenzie,BartoKozikLics10}.
We will use the formulation by Siggers~\cite{Siggers}; it will be important for our purposes
that Siggers polymorphisms have a fixed arity six and so for example on a fixed finite domain $D$ there is only a finite number of them.

\ifTR
\begin{definition} An operation $s\colon D^6 \to D$ is called a Siggers operation on $D$ if for each $x,y \in D$ we have
\begin{align*}
s(x, x, x, x, y, y) &= s(x, y, x, y, x, x) \\
s(y, y, x, x, x, x) &= s(x, x, y, x, y, x) \\
s(x, x, x, x, x, x) &= x.
\end{align*}
\end{definition}
\else
\begin{definition} An operation $s\colon D^6 \to D$ is called a Siggers operation on $D$ if it is idempotent
(i.e.\ $s(x, x, x, x, x, x) = x$ for all $x\in D$), and also
$
s(x, x, x, x, y, y) = s(x, y, x, y, x, x)
$
and
$s(y, y, x, x, x, x) = s(x, x, y, x, y, x)$
for all $x,y \in D$.
\end{definition}
\fi

The conjecture is usually stated for {\em core} languages. To reduce the number of definitions,
we will give an alternative formulation that avoids cores.
For a language $\Gamma$ on $D$ and a domain $D'\subseteq D$ let $\Gamma[{D'}]$ be the language obtained from $\Gamma$
by restricting each function to the domain $D'$.

\begin{definition} \label{def:SiggersPair}
Tuple $(g,s)$ will be called a {\em Siggers pair on a domain $D$} if $g$ is a unary operation on $D$
satisfying $g\circ g=g$
and $s$ is a Siggers operation on $g(D)\subseteq D$. 
We say that a crisp language $\Gamma$ on domain $D$ admits $(g,s)$
if $g$ is a unary polymorphism of $\Gamma$ and $s$ is a 6-ary polymorphism of $\Gamma[g(D)]$.
\end{definition}

\begin{theorem}[\cite{Siggers}]\label{nosiggershard} 
A crisp constraint language $\Gamma$ that does not admit a Siggers pair is NP-Hard.
\end{theorem}


\begin{conjecture}[A version of the Algebraic Dichotomy Conjecture] If a crisp language $\Gamma$ admits a Siggers pair, then $\CSP\Gamma$ is tractable.
\end{conjecture}

There has been remarkable progress on this conjecture. It has been verified for domains of size 2 \cite{schaefer78:complexity} and 3 \cite{bulatov06:3-elementjacm}, or for languages containing all unary relations on $D$ \cite{bulatov11:conservative}. It has also been shown that it is equivalent to its restriction for directed graphs (that is when $\Gamma$ contains a single binary relation $\varrho$) \cite{DBLP:conf/cp/BulinDJN13}. Further, the conjecture holds if $\varrho$ corresponds to a directed graph with no sources and sinks \cite{barto09:siam}.
 Nevertheless, in the general case the conjecture remains open.

\begin{definition} A crisp language $\Gamma$ is called a \emph{BJK} language if it satisfies one of the following:
\begin{itemize}
\item $CSP(\Gamma)$ is tractable
\item $\Gamma$ does not admit a Siggers pair.
\end{itemize}
\end{definition}

\begin{conjecture}[Another version of the Algebraic Dichotomy Conjecture] Every crisp language $\Gamma$ is a BJK language.
\end{conjecture}

\subsection{Hybrid (V)CSP setting}

\begin{definition} Let us call a family $\mathcal{H}$ of relational structures with a common signature 
 a {\emph structural restriction}. If all the relations in $\mathcal{H}$ are unary, we call $\mathcal{H}$ \emph{all-unary}.
\end{definition}

\begin{definition}[Hybrid CSP] Let $D$ be a finite domain, $\bGamma$ a constraint language over $D$, and $\mathcal{H}$ a structural restriction of the same signature as $\bGamma$. We define $\cspo_\mathcal{H}(\bGamma)$ as the following decision problem: given a relational structure $\mathbf{R} \in \mathcal{H}$ as input,  decide whether $\mathbf{R} \to \mathbf{\Gamma}$.
\end{definition}

\begin{definition}[Hybrid VCSP] Let $D$ be a finite domain, $\bGamma = (D,f_1, \dots, f_k)$ a valued constraint language over $D$, and $\mathcal{H}$ a structural restriction of the same signature as $\bGamma$.  We define $\vcspo_\mathcal{H}(\bGamma)$ as the class of instances of the following form.

An instance is a function from $D^V$ to $\Qc$ given by 
\begin{equation}\label{eq:hybridVCSPinst}
f_{\calI}(h)=\sum_{(f, \bv)\in T}{w(f,\bv)f(h(\bv))},
\end{equation}
where $V$ is a finite set of variables, $w(f,\bv)$ are positive numbers and $T$ is a finite set of constraints
 determined by some relational structure $\bR=(V,r_1,\ldots,r_k)\in\calH$ as follows:
$T=\{(f_i,\bv)\:|\:i\in[k],\bv\in r_i\}$.
The goal is to find an {\em assignment} (or {\em labeling}) $h\in D^V$ that minimizes $f_\calI$.
\label{def:hybridVCSPinst}
\end{definition}

\begin{definition} A structural restriction $\mathcal{H}$ is called \emph{effective} for a class of (valued) languages $\mathcal{C}$ if there is a language $\bGamma$ with $\Gamma \in \mathcal{C}$, of the same signature as $\mathcal{H}$, such that $\textsc{(V)CSP}(\Gamma)$ is NP-Hard, 
whereas $\textsc{(V)CSP}_{\mathcal{H}}(\bGamma)$ is tractable. 

$\calH$ is called {\em ineffective} for $\calC$ if 
for every $\bGamma$ with $\Gamma\in\calC$, of the same signature as $\mathcal{H}$,
 $\textsc{(V)CSP}(\Gamma)$
and $\textsc{(V)CSP}_{\mathcal{H}}(\bGamma)$ are either both tractable or both NP-hard.
\end{definition}
Note, some structural restrictions could potentially be neither effective nor ineffective for a given $\calC$
(since there exist intermediate complexity classes between NP-hard and tractable problems).

\begin{example} Let us give some examples of effective restrictions for the class $\mathcal{C}$ of all crisp languages.

Let $\calH$ be the set of $k$-colorable graphs for $k > 2$. Note that $k$-colorable graphs are exactly those that map homomorphically to the complete graph $K_k$. Therefore for the language $\Gamma = \{ \neq_D \}$ on domain $D$ with $|D| > 2$, we get that $\cspo_\calH(\bGamma)$ is tractable (with a constant time algorithm that outputs {\tt YES}), whereas $\CSP\Gamma$ is NP-Hard.

Similarly, also restricting to the class of planar graphs or perfect graphs is effective, since planar graphs are 4-colorable \cite{appel1977},
and for perfect graphs the {\sc Graph $k$-Colouring} problem is known to be solvable in polynomial time~\cite{Grotschel88:geometric}.

\end{example}

\section{Our Results}\label{sec:result}

Most of our results will apply to structural restrictions $\mathcal{H}$ that are {\em up-closed}. 
\begin{definition} A family of relational structures $\mathcal{H}$ is called {\em closed under inverse homomorphisms} (or {\em up-closed} for short) if whenever $\mathbf{R}' \to \mathbf{R}$ and $\mathbf{R} \in \mathcal{H}$, then also $\mathbf{R}' \in \mathcal{H}$.
\end{definition}

As examples of up-closed relational structures, let us mention directed acyclic graphs or $k$-colorable graphs. The proofs are straightforward. On the other hand, many natural graph classes do not possess this property, e.g.\ planar graphs and perfect graphs.

We introduce a notion of a chromatic number of relational structures that generalizes the usual chromatic number of graphs.

\begin{definition} Let $\mathbf{R} = (V, r_1, \dots, r_k)$ be a relational structure. A coloring of $\mathbf{R}$, that is a mapping $c \colon V \to [m]$, is \emph{improper} if there is a color $j \in [m]$ such that for each $i \in [k]$, the relation $r_i$ contains a monochromatic tuple of the color $j$. A coloring that is not improper is called \emph{proper}.

We define the \emph{chromatic number} $\chi(\mathbf{R})$ of $\mathbf{R}$ to be the smallest number of colors that can yield a proper coloring of $\mathbf{R}$. (If no proper coloring exists, we set $\chi(\mathbf{R})=\infty$; this will happen if e.g.\ $\mathbf{R}$ contains only one unary relation).
 Also, we define the \emph{ chromatic number} $\chi(\mathcal{H})$ of a structural restriction as
$$\chi(\mathcal{H}) = \sup \{ \chi(\mathbf{R}) : \mathbf{R} \in \mathcal{H} \}.$$

\end{definition}

\begin{theorem}\label{coloreffective} A structural restriction $\mathcal{H}$ with $\chi(\mathcal{H}) < \infty$ that is not all-unary is effective for the class of BJK languages.
\end{theorem}

\begin{theorem}\label{bjkineffective} An up-closed structural restriction $\mathcal{H}$ with $\chi(\mathcal{H}) = \infty$ is ineffective for the class of BJK languages. 

\end{theorem}

In particular, Theorem~\ref{bjkineffective} means that the Algebraic Dichotomy Conjecture would imply
that up-closed structural restrictions $\mathcal{H}$ with $\chi(\mathcal{H}) = \infty$ are ineffective for the class of all CSP languages.
Next, we state our results for valued languages. 

\begin{definition} A valued language is called \emph{conservative} if it contains all unary $\{0,1\}$-valued cost functions.
\end{definition}

\begin{definition} We say that a relational structure $\mathcal{H}$ \emph{does not restrict unaries} if for each $\mathbf{R} \in \mathcal{H}$
of the form $\mathbf{R} = (V, r_1, \dots, r_{i-1}, r_i, r_{i+1}, \dots, r_{k})$ with $\ar(r_i)=1$ and for each unary relation $r'_i\subseteq V$, we have
 $\mathbf{R}' \in \mathcal{H}$, where $\mathbf{R}' = (V, r_1, \dots, r_{i-1}, r'_i, r_{i+1}, \dots, r_{k})$.
\end{definition}

\begin{theorem}\label{vcspconsineff} An up-closed structural restriction $\calH$ with $\chi(\mathcal{H}) = \infty$ that does not restrict unaries is ineffective for the class of conservative valued  languages.
\end{theorem}

\ifTR
\begin{remark}
Note that our current techniques do not easily extend to other classes of VCSPs, e.g.\ finite-valued languages~\cite{tz13:stoc}.
Informally, the difficulty can be attributed to the fact that tractable finite-valued languages are characterized
by fractional polymorphisms {\em with an arbitrarily large support} (if the size of the domain is not fixed),
whereas for conservative languages we need two fractional polymorphisms that contain a constant number of operations
in the support, namely 2 and 3~\cite{kz13:jacm}.
\end{remark}
\fi

\ifTR
The proofs of the main theorems are described in the later sections. 
But first in Sec.~\ref{sec:OrderedCSP}-\ref{sec:MaxIndependentSet} we will list three implications of our theorems.
\else
Below we list three implications of our theorems. All missing proofs can be found in the full version
of the paper~\cite{KRT:arxiv15}.
\fi

\subsection{Ordered CSP}\label{sec:OrderedCSP}

One natural structural restriction to fixed template CSP is to introduce ordering of variables and request the constraints to respect the ordering.

\begin{definition} We call a relational structure $(V, r_1, \dots, r_k)$ \emph{ordered} if, after some identification of $V$ with $[n]$ for $n =|V|$, whenever $(v_1, \dots, v_{\ar(r_j)}) \in r_j$ for some $j = 1, \dots, k$, then $v_1 < \dots < v_{\ar(r_j)}$.
\end{definition}

\begin{theorem} Let $\mathcal{H}$ be the set of all ordered relational structures of some fixed signature.
Such structural restriction  $\mathcal{H}$ is ineffective for BJK languages and for conservative valued languages.
\end{theorem}
\begin{proof} It suffices to show that preconditions of Theorems \ref{bjkineffective} and \ref{vcspconsineff} hold.
\begin{itemize}
\item $\mathcal{H}$ up-closed:

Let $\mathbf{R} = (V, r_1, \dots, r_k) \in \mathcal{H}$ and $\mathbf{R}' \to \mathbf{R}$ where $\mathbf{R}' = (V', r'_1, \dots, r'_k)$ and let $h \colon V' \to V$ be the homomorphism. We may assume $V = [n]$. Let us define a partial order on $V'$ such that $v'_1 < v'_2$, if $h(v'_1) < h(v'_2)$. Extend this partial order arbitrarily to a total order on $[m]$ for $m = |V'|$ and identify $V'$ and $[m]$.

Now take $(v_1, \dots, v_{\ar(r'_j)}) \in r'_j$ for some $j = 1, \dots, k$ and since $(h(v_1), \dots, h(v_{\ar(r'_j)})) \in r_j$, we have $h(v_1) < \dots < h(v_{\ar(r'_j)})$ and thus also $v_1 < \dots < v_{\ar(r'_j)}$.

We have just verified that $\mathbf{R}' \in \mathcal{H}$.

\item $\chi(\mathcal{H}) = \infty$:

Fix $n \in \mathbb{N}$. We will construct $\mathbf{R} \in \mathcal{H}$ that cannot be properly colored with $n$ colors. Let $m$ be the maximal arity of relations in $\mathcal{H}$. Let $\mathbf{R} = (V, r_1, \dots, r_k)$ where $V = [n(m-1)+1]$ and for $j = 1, \dots, k$ we set $(v_1, \dots, v_{\ar(r_j)}) \in r_j$ if and only if $v_1 < \dots < v_{\ar(r_j)}$.

Clearly, $\mathbf{R} \in \mathcal{H}$. Now for any coloring with $n$ colors some color (say red) appears at least $m$ times. Let $v_1 < \dots < v_m$ be red elements of $V$. But then the tuples $(v_1, \dots, v_{\ar(r_j)})$ are red for $j = 1, \dots, k$ and hence the coloring is improper. 
\item $\calH$ does not restrict unaries: this follows directly from the definitions.
\end{itemize}
\end{proof}

This has an interesting consequence for graph homomorphism problems. Namely, restricting the input to directed acyclic graphs does not, assuming algebraic dichotomy conjecture, change the complexity of the problem.

\begin{corollary} For the class of directed acyclic graphs $\mathcal{H}$, algebraic dichotomy conjecture implies that for every language $\bGamma=(D,\varrho)$ with a binary relation $\varrho$, $\CSP\Gamma$ is tractable if and only if $\cspo_{\calH}(\bGamma)$ is tractable.
\end{corollary}

\begin{remark} A related result appeared in \cite{feder98:monotone}. The authors showed that a dichotomy for CSPs with input structures restricted to partial orders gives the dichotomy for all CSPs. However, no connection is shown between $\cspo_{\calH}(\bGamma)$ and $\CSP\Gamma$ as it is in our case.
\end{remark}


\subsection{Minor-closed families of graphs}\label{sec:MinorClosed}

\begin{theorem} Let the structural restriction $\mathcal{H}$ be a family of directed graphs such that the underlying family of undirected graphs is minor-closed. Then $\mathcal{H}$ is effective for BJK languages if and only if $\chi(\mathcal{H}) < \infty$.
\end{theorem}
\begin{proof} We use a result formulated as Lemma 2 in \cite{nesetril:minorclosed}, that relies on an old theorem by Mader \cite{mader67}: a minor-closed family of undirected graphs has either bounded chromatic number or contains all graphs.

In the first case we also have $\chi(\mathcal{H}) < \infty$ and $\mathcal{H}$ is effective by Theorem \ref{coloreffective}. In the other case, for each $G = (V, E) \in \mathcal{H}$ and each pair $x,y \in V$, $x \neq y$, we have $(x,y) \in E$ or $(y,x) \in E$. 

We will show that then $\mathcal{H}$ contains all directed acyclic graphs and thus is ineffective due to Theorem \ref{bjkineffective}. In fact, it suffices to show that $\mathcal{H}$ contains a total order (a complete directed acyclic graphs) of every size, since every directed graph is a minor of some total order.

To this end, fix $n \in \mathbb{N}$ and pick $G \in \mathcal{H}$ with $R(n,n)$ vertices, where $R(n,n)$ is the corresponding Ramsey number. We set $V(G) = [R(n,n)]$ and color an edge $(x,y)$ blue if $x > y$ and red if $x < y$. By Ramsey's Theorem we are guaranteed to find a monochromatic clique of size $n$. This clique is a minor of $G$ and gives us the desired total order.

\end{proof}

\subsection{Maximum Independent Set}\label{sec:MaxIndependentSet}

Although Theorem~\ref{vcspconsineff} is formulated for conservative languages, it also gives implications
for some optimization problems corresponding to non-conservative languages.
In this subsection, we will show that the classical problem of {\sc max weight independent set} is still intractable on some classes of graphs.

Given a class of undirected graphs $\calG$, we write  $\mwis_\calG$ to denote {\sc max weight independent set} 
problem
(with positive node weights) restricted to class $\calG$.
If $\calG$ is the class of all undirected graphs, let us write $\mwis$ instead of $\mwis_\calG$.
We say that $\calG$ is {\em up-closed} if it satisfies the following condition:
if $G,G'$ are undirected graphs such that $G\in\calG$ and $G'$ maps homomorphically to $G$,
then $G'\in\calG$.

\begin{theorem} Let $\calG$ be an up-closed family of undirected graphs with $\chi(\calG) = \infty$. Then $\mwis_\calG$ is NP-hard.
\label{th:MWIS}
\end{theorem}

To prove this theorem, consider language $\bGamma=(D,f,f_1,\ldots,f_k)$
where $D=\{0,1\}$, $f$ is the binary function with 
 $f(1,1) = \infty$ and $f(0,0) = f(0,1) = f(1,0) = 0$, and $\{f_1,\ldots,f_k\}$ is the set of all  $\{0,1\}$-valued unary functions on $D$.
Given a class of graphs $\calG$, we define a structural restriction $\calH(\calG)$ of the same signature as $\bGamma$ that does not restrict unaries as follows:
$$\calH(\calG) = \{(V, G, V_1, \dots, V_{k}) \colon G \in \vec\calG, V_1, \dots, V_{k} \subseteq V  = V(G) \}$$
where $\vec\calG$ denotes the family of all directed graphs 
that can be obtained by taking a graph $G\in\calG$ and orienting edges in an arbitrary way.




\begin{proposition} Let $\calG$ be a family of undirected graphs closed under taking induced subgraphs.
Then $\mwis_\calG$ and $\vcspo_{\calH(\calG)}(\bGamma)$ are polynomial-time equivalent.
\label{prop:MWIS:equiv}
\end{proposition}
\begin{proof} It will be convenient to treat $\mwis_\calG$ as the {\sc min weight independent set}, where the weight of each node is a negative rational number.
Clearly, this is equivalent to the original definition of $\mwis_\calG$. 

In one direction the reduction is trivial: any instance of $\mwis_\calG$ can be easily cast as an instance of $\vcspo_{\calH(\calG)}(\bGamma)$
(assuming that vertices labeled with 1 correspond to vertices of an independent set). 
Let us consider the other direction. Let $\calI$ be an instance of $\vcspo_{\calH(\calG)}(\bGamma)$.
Let $G=(V,E)\in\vec\calG$ be the corresponding graph.
After merging unary terms we can rewrite the objective function of $\calI$ as
$$
f_\calI(h)=\sum_{(u,v)\in E}w(u,v)f(h(u),h(v)) + \sum_{v\in V} w_v h(u) + {\mbox{\em const}}\qquad \forall h:V\rightarrow\{0,1\}
$$
where weights $w(u,v)$ are positive. 
Now set $V^- = \{ v \in V \colon w_v < 0 \}$ and let $G^-$ be the induced subgraph of $G$ on the vertex set $V^-$. Note that $G^- \in \vec\calG$. Now solve the min weight independent set problem on $G^-$ and label the chosen vertices with 1 and all others with 0. It is easy to see that this is an optimal assignment for $\calI$.

\end{proof}
We can now prove Theorem~\ref{th:MWIS}. It can be checked that any up-closed class of graphs is closed under taking induced subgraphs,
and so the precondition of Proposition~\ref{prop:MWIS:equiv} holds.
Problem $\vcspo(\Gamma)$ is polynomial-time equivalent to $\mwis$ and thus is NP-hard.
It is easy to check that up-closedness of $\calG$ implies up-closedness of $\calH(\calG)$.
Therefore, by Theorem~\ref{vcspconsineff} $\vcspo_{\calH(\calG)}(\bGamma)$ is also
NP-hard, and thus so is  $\mwis_\calG$ by Proposition~\ref{prop:MWIS:equiv}.

\begin{remark} Let us mention that up-closed graph classes $\calH$ with $\chi(\calH) = \infty$ can be non-trivial. Let $\calH^{odd}_k$ be the class of graphs with odd girth at least $k$. Then for example $\calH^{odd}_4$ is the class of triangle-free graphs. It can be checked that $\calH^{odd}_k$ is up-closed for every $k$ (homomorphic image of an odd cycle contains an odd cycle of equal or smaller length). For the unbounded chromatic number we refer to a classical result \cite{Erdös1966149} that states that the family of graphs $\calH_k$ with girth at least $k$ has $\chi(\calH_k) = \infty$. Since $\calH_k \subseteq \calH^{odd}_k$, we get also $\chi(\calH^{odd}_k) = \infty$.

\end{remark}

%

\section{Proof of Theorem \ref{coloreffective}}\label{sec:effproof}

We need to construct a BJK language $\bGamma$ such that $\CSP\Gamma$ is NP-Hard whereas $\cspo_\mathcal{H}(\bGamma)$ is polynomially tractable.

Let $n_1, \dots, n_k$ be the arities of relational structures in $\mathcal{H}$. Also take $m$ such that $m \ge \chi(\mathcal{H})$ and $m > 2$. We will define 
$\bGamma = (D,\varrho_1, \dots, \varrho_k)$ on the domain $D = D_1 \cup \dots \cup D_k$, where $D_i$ are pairwise disjoint copies of $[m]$. Also, let $d_i \colon [m] \to D$ such that $d_i(j)$ is the copy of $j$ in $D_i$. For $i \in [k]$ we set $$X^i_\infty = \{(a,  \dots,  a) \colon (a,  \dots,  a) \subseteq D^{n_i}, a \in D_i \} $$
and then define $\varrho_i = D^{n_i} \setminus X^i_\infty$.

\begin{itemize}
\item {\bf Hardness of $\CSP\Gamma$}: We will show that $\Gamma$ can express a certain coloring relation. Let us define a binary relation $\varrho \in \langle \Gamma \rangle$ as
$$
\varrho(x,y) = \left( \bigwedge_{i \in I_1} \left( \varrho_i(x) \wedge \varrho_i(y)\right) \right) \wedge  \left( \bigwedge_{i \in I_{\geq2}} \varrho_i(x,y,y, \dots y) \right)
$$
where $I_1$ is the set of indices of the unary relations in $\Gamma$ and $I_{\geq2}$ are the indices of the non-unary relations. Note that $I_{\geq2}$ is nonempty. Let
%
%
$D'=\bigcup_{i \in I_{\geq2}} D_i$, then  $|D'| \geq m$.
It can be checked that $(x,y) \notin \varrho$ if $x \notin D'$ or $y \notin D'$. Finally, for $x,y \in D'$, we clearly have $(x,y) \in \varrho$ if $x\neq y$ and also $(x,y) \notin \varrho$ for $x = y$, since for some $i \in I_{\geq2}$ we have $(x,\dots, x ) \in X^i_\infty$. That is, $\varrho(x,y)$ corresponds to a $\neq$ relation on $D'$ (which corresponds to $|D'|$-coloring) and since $|D'| \geq m > 2$ and $\varrho \in \langle \Gamma \rangle$
this makes $\CSP\Gamma$ NP-Hard by Lemma \ref{lem:exppower}(c).

\medskip
\item { \bf Tractability of $\cspo_\mathcal{H}(\bGamma)$}: We claim that a constant-time algorithm that outputs {\tt YES} is correct for every instance of $\cspo_\mathcal{H}(\bGamma)$.

Consider an instance given by a relational structure $\mathbf{R} = (V, r_1, \dots, r_k) \in \mathcal{H}$. 
Since $\chi(\mathbf{R}) \le m$,
there exists a proper coloring of $\mathbf{R}$ with colors $1, \dots, m$. Now, as the coloring is proper, for each color class $j \in [m]$, there exists $i = i(j)$ such that the relation $r_i$ has no monochromatic tuple in the color $j$. Let us define a map $s\colon V \to D$. If $j$ is the color of $v \in V$, then let $s(v) = d_{i(j)}(j)$. We claim this assignment is feasible.

Indeed, suppose not, then there exist index $i\in[k]$ and a tuple $\bv=(v_1,\ldots,v_{n_k})\in r_i$ such that $s(\bv)\notin \varrho_i$.
Thus, $s(\bv)=(a,\ldots,a)$ for some $a\in D_i$.
This means that $v_1,\ldots,v_{n_k}$ have the same color $j$
and $a=d_{i(j)}(j)$. Condition $d_{i(j)}(j)\in D_i$ implies that $i(j)=i$.
We obtained that relation $r_i$ for $i=i(j)$ contains a monochromatic tuple in color $j$, which is a contradiction.


\end{itemize}

It remains to say that $\Gamma$ is a BJK language. First, observe that language $\{\varrho\}$ is a BJK
language (binary relation $\rho$ corresponds to a digraph without sources and sinks, for which
the Algebraic Dichotomy conjecture has been established in~\cite{barto09:siam}).
Since $\{\varrho\}$ is NP-hard, we obtain that $\rho$ does not admit a Siggers pair.
By Lemma~\ref{lemma:SiggersPairs'} below, $\Gamma$ also does not admit a Siggers operation, and thus is a BJK language.


\begin{lemma}
Let $\Gamma$ be a crisp language on a domain $D$ that admits a Siggers pair $(g,s)$ with $A\!=\!g(D)\!\subseteq\! D$. 
Then language $\langle\Gamma\rangle$ also admits the Siggers pair $(g,s)$. 
\label{lemma:SiggersPairs'}
\end{lemma}
\begin{proof}
We need to show the following for every crisp function $f\in\langle\Gamma\rangle$ of arity $k=\ar(f)$:
(i) $f$ admits $g$ as a unary polymorpshism; (ii) $f_{|A}$ (the restriction of $f$ to $A^{k}$) admits $s$ as a 6-ary polymorphism.
The first claim holds by Lemma~\ref{lem:exppower}(a). We will show that $f_{|A}\in\langle\Gamma[A]\rangle$,
then the second claim will again follow by Lemma~\ref{lem:exppower}(a).

Since $f\in\langle\Gamma\rangle$, there exists a $\Gamma$-instance $\calI$ with $n\ge k$ variables such that
$$
f(\bx)=\min_{\by\in D^{n-k}} f_\calI(\bx,\by) \qquad \forall \bx \in D^k
$$
Define a function $f':A^k\rightarrow\{0,\infty\}$ via
$$
f'(\bx)=\min_{\by\in A^{n-k}} f_\calI(\bx,\by) \qquad \forall \bx \in A^k
$$
By construction, $f'\in\langle\Gamma[A]\rangle$. It thus suffices to prove that $f_{|A}=f'$.
Consider $\bx\in A^k$. Clearly, we have $f(\bx)\le f'(\bx)$ (or equivalenty
$\bx\in \dom f'$ implies $\bx\in\dom f$). Suppose that $\bx\in \dom f$.
Then there exists $\by\in D^{n-k}$ such that $(\bx,\by)\in \dom f_{\calI}$.
Since $g$ is a polymorphism of $f_\calI$, we obtain $(g(\bx),g(\by))\in \dom f_{\calI}$.
The properties of $g$ stated in Definition~\ref{def:SiggersPair}, in particular the idempotence, and the fact $x\in A^k$
give that $g(\bx)=\bx$ and $g(\by)\in A^{n-k}$.
Therefore, $(\bx,\by')\in\dom f_{\calI}$ for some $\by'\in A^{n-k}$ and so $\bx\in\dom f'$.
\end{proof}

\section{Constructing a ``lifted'' language}\label{sec:construction}

For both Theorems \ref{bjkineffective} and \ref{vcspconsineff} we need to show that tractability of the restricted problem implies tractability of the unrestricted one. 

Let $\bGamma=(D,f_1,\ldots,f_k)$ be a language of the same signature as $\calH$
and $\bR$ be a relational structure in $\calH$.
In this section we will construct a language $\Gamma_\bR$ of finite size on a larger domain, based on $\bGamma$ and $\bR$.
Our strategy will then be to link languages $\{\Gamma_\bR\::\:\bR\in\calH\}$, in terms of tractability, to both $\vcspo_\mathcal{H}(\bGamma)$ and $\VCSP\Gamma$.
Namely, we will first prove  the following.

\begin{proposition}\label{vcspinherittract} Suppose that $\mathcal{H}$ is up-closed, $\bR \in \calH$ and $\bGamma$ is a (valued) language.
Then there is a polynomial-time reduction from $\textsc{(V)CSP}(\Gamma_\bR)$ to $\textsc{(V)CSP}_\mathcal{H}(\bGamma)$.
Consequently, \\
(a) 
if $\textsc{(V)CSP}_\mathcal{H}(\bGamma)$ is tractable, then so is $\textsc{(V)CSP}(\Gamma_\bR)$; \\
(b) if $\textsc{(V)CSP}(\Gamma_\bR)$ is NP-hard, then so is $\textsc{(V)CSP}_\mathcal{H}(\bGamma)$.
\end{proposition}


Using algebraic tools, we will then show in sections \ref{sec:bjk} and \ref{sec:vcspcons}  how tractability of
$\Gamma_\bR$ for all $\bR\in\calH$
 implies tractability of $\bGamma$ for $\bGamma$ lying in the particular language classes.

\subsection{Construction of $\Gamma_\bR$}\label{sec:construction'}
Let us fix a relational structure $\bR=(V,r_1,\ldots,r_k)$.
For each  $v \in V$ we create a unique copy of the domain $D$, and denote it $D_{v}$.
We then define
$$
D_\bR=\bigcup_{v\in V} D_v.
$$

For   $v \in V$ define a mapping $d_{v} \colon D \to D_\bR$ such that $d_{v}(a)$ is the copy of $a$ in $D_{v}$. Also for tuples $\mathbf{a} = (a_1, \dots, a_p) \in D^p$ and $\bv=(v_1,\ldots,v_p)\in V^p$ we set $d_{\mathbf{v}}(\mathbf{a}) = (d_{v_1}(a_1), \dots, d_{v_p}(a_p))$. 

For the opposite direction,
let  $d(b)$ for $b \in D_\bR$ be the natural projection of $b$ on $D$, and for a tuple $\mathbf{b} = (b_1, \dots, b_p)$ let $d(\mathbf{b}) = (d(b_1), \dots, d(b_p))$.

Now for a cost function $f \in \CostF_D$ and $\mathbf{v} \in V^{\ar(f)}$ we will define a cost function on $D_\bR$ of the same arity as $f$ via
$$f^{\bv}(\mathbf{x})  = \begin{cases}
               f(\mathbf{y}) \quad \text{if} \,\, \mathbf{x} = d_{\mathbf{v}}(\mathbf{y}) \,\, \text{for some}\,\, \mathbf{y} \in D^{\ar(f)} \\
               \infty \qquad \text{otherwise} \\
            \end{cases}
\qquad\forall \mathbf{x}\in D_\bR^{\ar(f)}
$$
Note that this equation is well-defined since the mapping $d_{\mathbf{v}}$ is injective. Furthermore, we have the following
properties.
\begin{lemma} \label{vcsptriv_lemma}
(a) $f(\mathbf{y}) = f^{\mathbf{v}}(d_{\mathbf{v}}(\mathbf{y}))$ for any $\mathbf{y} \in D^{\ar(f)}$.
(b) $f^{\mathbf{v}}(\mathbf{x}) = f(d(\mathbf{x}))$ for $\mathbf{x} \in \dom f^{\mathbf{v}}$.
\end{lemma}

Finally, we construct the sought language $\Gamma_\bR$ on domain $D_\bR$ as follows:
$$\Gamma_\bR = \{f_i^\bv \::\: i\in[k], \bv \in r_i \} \cup \{ D_v \::\: v \in V\} $$
where relation
$D_v\subseteq D_\bR$ is treated as a unary function $D_v:D_\bR\rightarrow\{0,\infty\}$.

\begin{remark}
We note that there are some parallels between the construction above and the notion of {\em multi-sorted relations}~\cite{Bulatov03:multi-sorted}.
Our approach, however, is different from that in~\cite{Bulatov03:multi-sorted}:
the language $\Gamma_\bR$ that we have constructed is a standard (non-multi-sorted) language, which allows us to apply many results known for \textsc{(V)CSP}s.
\end{remark}

\begin{remark}
Lifted language $\Gamma_\bR$ should not be confused with ``$G$-lifted languages''
used in~\cite{GreenCohen:AI08}; despite similar names, the constructions are not
related. 
\end{remark}


\subsection{Proof of Proposition \ref{vcspinherittract}}\label{sec:vcspinherittract}
Consider a $\Gamma_\bR$-instance $\calI$ with the set of variables $U$ and the objective function
\begin{equation*}
f_\calI(h)=\sum_{(f_i^\bv,\bu)\in T} w(f_i^\bv,\bu) f_i^\bv(h(\bu))
+\sum_{(D_v,u)\in T'} w(D_v,u) D_v(h(u))
\qquad\forall h:U\rightarrow D_\bR
\end{equation*}
We can assume w.l.o.g.\ that each variable $u\in U$ is involved in at least one constraint of arity 2 or higher. 
(If $u$ is involved in only unary constraints, we can find an optimal solution $h(u)$ independently of other
variables, and then remove $u$.) 
By construction, each constraint induced by a cost function in $\Gamma_\bR$ restricts each of its variables to a particular copy of $D$ in $D_\bR$. 
If different constraints restrict the same variable $u\in U$ to different copies of $D$, then clearly $\calI$ has no feasible solutions;
we then say that $\calI$ is {\em trivially infeasible}. Note that we can test this in polynomial time.

Now suppose that $\calI$ is not trivially infeasible. Then for each $u\in U$ we can determine in polynomial time 
 node $v\in V$ such that all constraints in $\calI$ that involve $u$ restrict solution $h_u$ to $D_{v}$. Let $\varphi:U\rightarrow V$
be the corresponding mapping that gives $v=\varphi(u)$. We then have the following property:
\begin{proposition}
If $(f_i^\bv,\bu)\in T$, then $\bv=\varphi(\bu)$, where $\varphi$ is applied component-wise.
\label{prop:GAJKHFKAJSHGKA}
\end{proposition}
\begin{proof}
Let $\bu=(u_1,\ldots,u_p)$ and $\bv=(v_1,\ldots,v_p)$.
We assumed that the constraint $(f_i^\bv,\bu)$ restricts variable $h(u_j)$ to the domain $D_{\varphi(u_j)}$. By definition of $f_i^\bv$, this function
restricts its $j$-th argument to the domain $D_{v_j}$. Thus, $\varphi(u_j)=v_j$.
\end{proof}

Consider an instance $\tilde\calI$ with the set of variables $U$, the domain $D$ and the cost function
\begin{equation*}
 f_{\tilde\calI}(\tilde h)=\sum_{(f_i^\bv,\bu)\in T} w(f_i^\bv,\bu)f_i(\tilde h(\bu))\qquad\forall \tilde h:U\rightarrow D.
\end{equation*}
We claim that solving the instance $\calI$ is equivalent to solving the instance $\tilde\calI$.
Indeed, let $\calS$ be the set of assignments $h:U\rightarrow D_\bR$ that are not ``trivially infeasible'' for $\calI$, i.e.\ that satisfy $h(u)\in D_{\varphi(u)}$ for all $u\in U$.
Let $\tilde\calS$ be the set of assignments $\tilde h:U\rightarrow D$.
It can be seen that $f_{\calI}(h)=\infty$ if $h\notin\calS$, and there is a cost-preserving bijection $\tilde\calS\rightarrow \calS$ that maps assignment $\tilde h\in\tilde S$
to the assignment defined by $h\in\calS$ with $h(u)=d_{\varphi(u)}(\tilde h(u))$. This implies the claim.

We will show next that $\tilde\calI\in \textsc{(V)CSP}_\mathcal{H}(\bGamma)$; this will imply the claim of Proposition~\ref{vcspinherittract}.

Define relational structure $\tilde\bR=(U,\tilde r_1,\ldots,\tilde r_k)$ as follows:
%
$\tilde r_i=\{\bu\:|\:(f_i^\bv,\bu)\in T\}$ for $i\in[k]$.
It defines the set of constraints $\tilde T=\{(f_i,\bu)\:|\:i\in[k],\bu\in \tilde r_i\}=\{(f_i,\bu)\:|\:i\in[k],(f_i^\bv,\bu)\in T\}$.
Using Proposition~\ref{prop:GAJKHFKAJSHGKA}, it can be checked that there is a natural isomorphism between $T$ and $\tilde T$,
and $\tilde T$ defines the set of constraints for the instance $\tilde\calI$ as in Definition~\ref{def:hybridVCSPinst}.
It thus suffices to prove that $\tilde\bR\in\calH$.

We claim that the mapping $\varphi$ is a homomorphism from $\tilde\bR$ to $\bR$.
Indeed, we need to show that if $\bu\in \tilde r_i$, then $\varphi(\bu)\in r_i$.
We have $(f_i^\bv,\bu)\in T$ where $\bv=\varphi(\bu)$ by Proposition~\ref{prop:GAJKHFKAJSHGKA}.
The condition $f_i^\bv\in\Gamma_\bR$ implies that $\bv\in r_i$, or equivalently $\varphi(\bu)\in r_i$.

We showed that $\tilde\bR \stackrel{\varphi}{\to} \bR$. Since $\bR\in\calH$ and $\calH$ is up-closed, we obtain that $\tilde\bR\in\calH$, as desired.

\section{Proof of Theorem \ref{bjkineffective}}\label{sec:bjk}


We will show the following result.

\begin{proposition}\label{inherit_siggers} Let $\mathcal{H}$ be a structural restriction with $\chi(\mathcal{H}) = \infty$ and $\bGamma$ a constraint language of the same signature as $\calH$. If for every $\bR \in \calH$ language $\Gamma_\bR$ admits a Siggers pair $(g_\bR,s_\bR)$, then $\Gamma$
also admits Siggers pair.
\end{proposition}
Before giving a proof, we describe how this proposition implies Theorem~\ref{bjkineffective}.
First, suppose that $\cspo_\calH(\bGamma)$ is tractable. Proposition~\ref{vcspinherittract}(a)
gives that  for every $\bR\in\calH$ language $\Gamma_\bR$ is tractable, and thus admits a Siggers pair
by Theorem~\ref{nosiggershard}. Proposition~\ref{inherit_siggers} then gives
that $\Gamma$ also admits a Siggers pair, and thus is tractable since $\Gamma$ is a BJK language.

Now suppose that $\cspo_\calH(\bGamma)$ is not tractable. Then $\Gamma$ is also not tractable, and thus does
not admit a Siggers pair (since  $\Gamma$  is a BJK language).
By Proposition~\ref{inherit_siggers}, there exists $\bR\in\calH$ such that the language $\Gamma_\bR$
does not admit any Siggers pair, and thus is NP-hard by Theorem~\ref{nosiggershard}.
Proposition~\ref{vcspinherittract}(b) now gives that $\cspo_\calH(\bGamma)$ is NP-hard.
This also implies NP-hardness of $\CSP\Gamma$.

\begin{proof}[Proof (of Proposition~\ref{inherit_siggers}).]
Let $\mathcal{S}$ be the (finite!) set of Siggers pairs $(g,s)$ on the domain $D$. Choose $\bR \in \calH$
such that 
 $\chi(\mathbf{R}) > |\mathcal S|$. Let $(g,s)$ be a Siggers pair admitted by $\Gamma_\bR$. 
We will use the notation from Sec.~\ref{sec:construction'} for the chosen $\bR=(V,r_1,\ldots,r_k)$.
Note that $g$ is a unary operation on $D_\bR$ and $s$ is a 6-ary operation on $A$,
where we denoted $A=g(D_\bR)\subseteq D_\bR$.

For each $v \in V$, we  denote the restriction $g_{|D_v}$ simply as $g_v$.
The fact $D_v\in\Gamma_\bR$ 
gives that $g_v(D_v)\subseteq D_v$. Since this holds for each $v\in V$
and $D_\bR$ is a disjoint union of $\{D_v\::\:v\in V\}$, we obtain that $g_v(D_v)=D_v\cap g(D_\bR)$.
We denote $A_v=g_v(D_v)=D_v\cap A$.

Similarly, we denote the restriction $s_{|A_v}$ simply as $s_v$.
We have $D_v\in\Gamma_\bR$ and so $A_v=D_v\cap A\in\Gamma_\bR[A]$.
Operation $s$ is a polymorphism of $\Gamma_\bR[A]$, therefore
 $s_v(A_v,\ldots,A_v)\subseteq A_v$.
This shows that
$(g_v,s_v)$ is a Siggers pair on $D_v$.
It can also be identified with a Siggers pair on domain $D$
(via a natural isomorphism $\sim_v$ induced by the bijection $d:D_v\rightarrow D$), and hence we can write $(g_v,s_v) \in \mathcal{S}$. 

We use the pairs $\{(g_v,s_v)\}_{v \in V}$ to color the elements of $V$. Since $\chi(\mathbf{R}) > |\mathcal{S}|$, this coloring is improper and therefore there is a Siggers pair on $D$
(``color'') $(\tilde g,\tilde s) \in \mathcal{S}$ and
tuples $\mathbf{v}^i = (v^i_1, \ldots, v^i_{\ar(r_i)}) \in r_i$  for each $i\in[k]$ such that $(g_v,s_v) \sim_v  (\tilde g,\tilde s)$ for all $v \in \{v^i_j \colon i \in [k], j \in [\ar(r_i)]  \}$. 
We will show next that $\Gamma$ admits $(\tilde g,\tilde s)$. We denote $\tilde A=\tilde g(D)\subseteq D$, then $\tilde s$ is a Siggers operation on $\tilde A$.

Consider index $i\in[k]$, and let $p$ be the arity of $f_i$. We need to show two facts.
\begin{itemize}
\item  $\tilde g$ preserves $f_i$.
Consider vector $\bx=(x_1,\ldots,x_p)\in \dom f_i$.
First, we realize that 
$$\tilde g(\mathbf{x}) 
= d(g_{v^i_1}(d_{v^i_1}(x_1)), \ldots, g_{v^i_p}(d_{v^i_p}(x_p))) = d(g({\mathbf{y}}))$$
for some ${\mathbf{y}} \in \left(D_\bR\right)^p$ 
(namely, ${\mathbf{y}}=d_{\bv^i}(\mathbf{x})$). 
Since $\mathbf{x} \in \dom f_i$, we also have that $\mathbf{y} \in \dom f_i^{\mathbf{v}^i}$.
As $g$ is a polymorphism of $\Gamma_\bR$, we get that $g({\mathbf{y}}) \in \dom f_i^{\mathbf{v}^i}$. But this gives $d(g({\mathbf{y}})) \in \dom f_i$ and we may conclude the proof.
\item  $\tilde s$ preserves $(f_i)_{|\tilde A}$ (which is the restriction of $f_i$ to $\tilde A$).
Let $\bx$ be a matrix with $p$ columns (denoted as $\mathbf{x}_1, \ldots, \mathbf{x}_p$)
and $6$ rows (denoted as $\mathbf{x}^1, \ldots, \mathbf{x}^6$) such that $\mathbf{x}^1, \ldots, \mathbf{x}^6 \in [\dom f_i]\cap \tilde A^p$. First, we realize that 
$$\tilde s(\mathbf{x}^1, \ldots, \mathbf{x}^6) = (\tilde s(\mathbf{x}_1), \ldots, \tilde s(\mathbf{x}_p)) 
= d(s_{v^i_1}(d_{v^i_1}(\mathbf{x}_1)), \ldots, s_{v^i_p}(d_{v^i_p}(\mathbf{x}_p))) = d(s({\mathbf{y}}^1, \ldots, {\mathbf{y}}^6))$$
for some ${\mathbf{y}}^1, \ldots, {\mathbf{y}}^6 \in \left(D_\bR\right)^p$ 
(namely, ${\mathbf{y}}^j=d_{\bv^i}(\mathbf{x}^j)$). 
Since $\mathbf{x}^1, \ldots, \mathbf{x}^6 \in [\dom f_i]\cap \tilde A^p$, we also have that $\mathbf{y}^1, \ldots, \mathbf{y}^6 \in [\dom f_i^{\mathbf{v}^i}]\cap A^p$.
As $s$ is a polymorphism of $\Gamma_\bR[A]$, we get that $s({\mathbf{y}}^1, \ldots, {\mathbf{y}}^6) \in \dom f_i^{\mathbf{v}^i}$. But this gives $d(s({\mathbf{y}}^1, \ldots, {\mathbf{y}}^6)) \in \dom f_i$ and we may conclude the proof.
\end{itemize}


\end{proof}

\section{Proof of Theorem \ref{vcspconsineff}}\label{sec:vcspcons}

For a relational structure $\bR=(V,r_1,\ldots,r_k)$ we define its {\em unary completion} $\bR'$ as follows:
take $\bR$ and replace every unary relation $r_i$ in $\bR$ with the unary relation $r'_i=V$.
Since $\calH$ does not restrict unaries, we have $\bR'\in\calH$ for each $\bR\in\calH$.
Let $\calH'\subseteq\calH$ be the set of unary-complete relational structures in $\calH$,
i.e.\ those structures $\bR\in\calH$ that satisfy $\bR'=\bR$. 
It follows from the definition that if a coloring $c:V\rightarrow[m]$ is improper for $\bR$,
then it is also improper for $\bR'$. Equivalently, if it is proper for $\bR'$, then it is proper for $\bR$.
Therefore, $\chi(\bR')\ge \chi(\bR)$, and consequently $\chi(\calH')=\infty$ (since $\chi(\calH)=\infty$).


\begin{proposition}\label{consinherittract} Suppose that $\bR$ is unary-complete relational structure and $\bGamma$ is
a conservative valued language. Then there is a polynomial-time reduction from $\vcsp{\Gamma_\bR\cup \Delta_\bR}$ to $\vcsp{\Gamma_\bR}$,
where $\Delta_\bR$ is the set of $\{0,1\}$-valued unary functions on the domain $D_\bR$.
Consequently, \\
(a) if $\vcsp{\Gamma_\bR}$ is tractable, then so is $\vcsp{\Gamma_\bR\cup \Delta_\bR}$; \\
(b) if $\vcsp{\Gamma_\bR\cup \Delta_\bR}$ is NP-hard, then so is $\vcsp{\Gamma_\bR}$.
\end{proposition}

\begin{proposition}\label{consijheritsiggers} Let $\mathcal{H}'$ be a structural restriction with $\chi(\mathcal{H}') = \infty$ and $\bGamma$ a conservative valued language. If  $\vcsp{\Gamma_\bR \cup \Delta_\bR}$ is tractable for every $\bR \in \calH'$, then $\vcsp\Gamma$ is tractable.
\end{proposition}

Let us describe how these propositions imply Theorem \ref{vcspinherittract}.
First, suppose that $\vcspo_\calH(\bGamma)$ is tractable.
Propositions~\ref{vcspinherittract}(a) and~\ref{consinherittract}(a)
give that for every $\bR\in\calH'$ language $\Gamma_\bR\cup\Delta_\bR$ is tractable.
Thus, by Proposition~\ref{consijheritsiggers} the language $\Gamma$ is tractable.

Now suppose that $\vcspo_\calH(\bGamma)$ is not tractable. Then $\Gamma$ is also not tractable.
By Proposition~\ref{consijheritsiggers} there exists $\bR\in\calH'$ such that $\Gamma_\bR\cup\Delta_\bR$
is not tractable. Language  $\Gamma_\bR\cup\Delta_\bR$ must then be NP-hard
(since it is conservative, and the dichotomy for conservative valued language has been established in \cite{kz13:jacm},
see Theorem~\ref{vcspconsthm} below).
By Propositions \ref{consinherittract}(b) and \ref{vcspinherittract}(b)
we obtain that $\vcspo_\calH(\bGamma)$ is NP-hard.
This also implies NP-hardness of $\vcsp\Gamma$.

\subsection{Proof of Proposition \ref{consinherittract}}

Consider an instance $\calI$ of $\Gamma_\bR\cup \Delta_\bR$ with the set of variables $U$ and the objective function
\begin{equation*}
f_\calI(h)=\sum_{(f,\bu)\in T} w(f,\bu) f(h(\bu)) 
+\sum_{(D_v,u)\in T'} w(D_v,u) D_v(h(\bu))
\qquad\forall h:U\rightarrow D_\bR
\end{equation*}
We can assume w.l.o.g.\ that each variable $u\in U$ is involved in at least one constraint of arity 2 or higher
(by the same argument as in Sec.~\ref{sec:vcspinherittract}).
Consider $u\in U$, then by the assumption there exists $v\in V$
such that $h(u)\in D_v$ for any feasible assignment $h$. (We use the notation from Sec.~\ref{sec:construction'}.)
Let us modify the instance $\calI$ by replacing each constraint of the form $(f,u)\in T,f\in\Delta_\bR$
with $(f',u)$, where $f':D_\bR\rightarrow \Qc$ is defined via
$$
f'(x)=\begin{cases}
f(x) & \mbox{if }x\in D_v \\
\infty & \mbox{otherwise}
\end{cases}
\qquad \forall x\in D_\bR
$$ 
Clearly, this transformation preserves optimal solutions of $\calI$. Using the definition of $\Gamma_\bR$
and the facts that $\bR$ is unary-complete and $\bGamma$ is conservative, we conclude that $f'\in \Gamma_\bR$.
After applying this transformation for all $u\in U$ with obtain an equivalent instance $\calI'\in\vcsp{\Gamma_\bR}$.
This implies the claim.

\ignore{
Let us take $n \in \mathbb{N}$. We aim to find $m \in \mathbb{N}$ such that for each instance $\mathcal{I}$ of $VCSP(\Gamma_\mathcal{H}^n \cup U_n)$ there is an instance $\mathcal{I}'$ of $VCSP(\Gamma_\mathcal{H}^m)$ such that solving $\mathcal{I}'$ to optimality implies solving $\mathcal{I}$ to optimality.

For $i = 1, \dots, n$, we consider a relational structure $\mathbf{R}'_i = (V_i, \bar{\varrho}^{\mathbf{R}_i}_1, \dots, \bar{\varrho}^{\mathbf{R}_i}_k)$ where for $j = 1, \dots, k$

$$\bar{\varrho}^{\mathbf{R}_i}_j = \begin{cases} 
\varrho^{\mathbf{R}_i}_j \quad &\text{if} \,\, \varrho^{\mathbf{R}_i}_j \, \text{is not unary} \\
V_i \quad & \text{otherwise}.
\end{cases}
$$

Since $\mathcal{H}$ does not restrict unaries, we have $\mathbf{R}'_i \in \mathcal{H}$ for each $i = 1, \dots, n$. We take $m \in \mathbb{N}$ such that $\{\mathbf{R}'_1, \dots, \mathbf{R}'_n \} \subseteq \{ \mathbf{R}_1, \dots, \mathbf{R}_m \}$.

Now take an instance $\mathcal{I}$ of $VCSP(\Gamma_\mathcal{H}^n \cup U_n)$ with variable set $U$ of and finite multiset of constraints $T = T_\Gamma \cup T_U$, where
$$T_\Gamma = \{ (f,\mathbf{u}) \colon (f,\mathbf{u}) \in T,\, f \in \Gamma_\mathcal{H}^n\} \qquad \text{and} \qquad T_U = \{ (f,\mathbf{u}) \colon (f,\mathbf{u}) \in T,\, f \in U_n\}.$$
Thus $\calI$ decomposes into 
$$\calI(s) = \mathcal{I}_\Gamma(s) + \mathcal{I}_U(s) = \sum_{(f,\mathbf{u}) \in T_\Gamma} f(s(\mathbf{u})) + \sum_{(f,\mathbf{u}) \in T_U} f(s(\mathbf{u})) $$
and the active variables $U' \subseteq U$ of $\mathcal{I}_\Gamma$ can be decomposed using Lemma \ref{lemma_decomp} as $U' = \bigcup_{i \leq n, v \in V_i} U^{\mathbf{R}_i}_v$. Variables in $U \setminus U'$ are constrained only by unary functions from $U_n$ and therefore they be efficiently minimized out. We may hence assume $U = U'$.

Now let us construct an instance $\mathcal{I}'$ of $VCSP(\Gamma_\mathcal{H}^m)$ on the variable set $U$ with the multiset of constraints $T' = T'_\Gamma \cup T'_U$ specified as follows

$$T'_\Gamma = \{((f_j)^{\mathbf{R}'_i}_\mathbf{v}, \mathbf{u}) \colon ((f_j)^{\mathbf{R}'_i}_\mathbf{v}, \mathbf{u}) \in T_\Gamma \} \qquad \text{and} \qquad \{ (f_{|D^{\mathbf{R}'_i}_v}, u ) \colon (f,u) \in T_U, \, u \in U^{\mathbf{R}_i}_v\}$$
where the ``restriction'' of $f$ is defined as
$$f_{|D^{\mathbf{R}'_i}_v} (u) = \begin{cases} f(u) \quad &\text{if} \,\, u \in D^{\mathbf{R}'_i}_v \\
\infty \quad& \text{otherwise}.
\end{cases}
$$
To make sure $\calI'$ is indeed an instance of $\Gamma^m_\calH$ we now show $f_{|D^{\mathbf{R}'_i}_v} \in \Gamma^m_\calH$ for every $f \in U_n$, $\mathbf{R}'_i \in \calH$, and $v \in V$.
First, consider $g \colon D \to \Qc$ defined as $g(a) = f_{|D^{\mathbf{R}'_i}_v}(d^{\mathbf{R}'_i}_v(a))$. This a $\{0,1\}$-valued function on domain $D$ and thus, as $\Gamma$ is conservative, we have $g = f_j$ for some $j = 1, \dots, k$ with $f_j \in \Gamma$. But then by construction of $\Gamma^m_\calH$ and due to the fact that $v \in \bar{\varrho}^{\mathbf{R}_i}_j = V_i$ we have that $(f_j)^{\mathbf{R}'_i}_v \in \Gamma^m_\calH$. Finally, it is immediate to check that $f_{|D^{\mathbf{R}'_i}_v} = (f_j)^{\mathbf{R}'_i}_v$.

The remaining step is to construct a cost-preserving bijection between feasible solution to $\calI$ and to $\calI'$. In fact, we will construct two cost-preserving injective mappings between the sets $\mathcal{S} = \{s \colon U \to D_n: f_{\mathcal{I}}(s) < \infty\}$ and $\mathcal{S}' = \{s' \colon U \to D_m: f_{\mathcal{I}'}(s) < \infty\}$. Namely, $h \colon \mathcal{S} \to \mathcal{S}'$ defined pointwise as $h(s)(u) =  d^{\mathbf{R}'_i}_v(d(s(u)))$ where $i$ and $v$ are such that $u \in D^{\mathbf{R}_i}_v$ and conversely $h' \colon \mathcal{S} \to \mathcal{S}'$ defined also pointwise as $h'(s)(u) = d^{\mathbf{R}_i}_v(d(s(u)))$ where $i$ and $v$ are such that $u \in D^{\mathbf{R}'_i}_v$. It is straightforward to check that both mappings well-defined and are injective. Preserving costs follows from Lemma \ref{vcsptriv_lemma}(a),(b) in a routine way.
}

\subsection{Proof of Proposition \ref{consijheritsiggers}}

First, we will recall the result on tractability of conservative VCSP languages from~\cite{kz13:jacm}.

\medskip

A subset $M \subseteq P$, where $P = \{ (a,b) \in D^2, a\neq b \}$ will be called \emph{symmetric} if $(a,b) \in M$ if and only if $(b,a) \in M$. Sometimes, we will abuse notation slightly by writing $\{a,b\} \in M$.

\begin{definition}\label{def:mm}
A fractional operation $\sigma = \frac12 \chi_\sqcap + \frac12 \chi_\sqcup$, where $\sqcap,\sqcup:D^2 \to D$, is called
a \emph{symmetric tournament pair} (STP) on symmetric $M \subseteq P$ if both operations $\sqcap,\sqcup$ are commutative on $M$, i.e.\ $a\sqcap b=b \sqcap a$ and $a\sqcup b=b \sqcup a$ for all $(a,b) \in M$, and
$(a\sqcap b,a\sqcup b)$ is a permutation of $(a,b)$ for all $(a,b) \in D^2$.
\end{definition}

\begin{definition} A fractional operation $\mu = \frac13 \chi_{F_1} + \frac13 \chi_{F_2} + \frac13 \chi_{F_3}$, where $F_1, F_2, F_3 \colon D^3 \to D$, is called
an \emph{MJN} on symmetric $M \subseteq P$ if $(F_1(a,b,c), F_2(a,b,c), F_3(a,b,c))$ is a permutation of $(a,b,c)$ for $a,b,c \in D$ and if whenever $\{a,b,c \} = \{x,y\}$ for some $\{x,y\} \in M$
then $F_1(a,b,c) = F_2(a,b,c)$ is the unique majority element among $a,b,c$ (that occurs twice) and $F_3(a,b,c)$ is the unique minority element among $a,b,c$ (that occurs once). 
\end{definition}

\begin{theorem}[\cite{kz13:jacm}]\label{vcspconsthm} Let $\Gamma$ be a conservative valued language and $P = \{(a,b) \in D^2, a\neq b \}$.
If there is a symmetric set $M'\subseteq P$ such that $\Gamma$ admits an STP on $M'$
and an MJN on $P \setminus M'$ as fractional polymorphisms, then $\vcspo(\Gamma)$ is tractable. Otherwise, $\vcspo(\Gamma)$ is
NP-hard.
\end{theorem}

Now we are ready to prove Proposition \ref{consijheritsiggers} and thus conclude the proof of Theorem \ref{vcspconsineff}.
\medskip

\begin{proof}

Let $\mathcal{S}$ be the (finite!) set

$$\{(\sigma,\mu,M) \colon \sigma \, \text{is STP on} \, M, \,\mu \, \text{is MJN on} \, P \setminus M, M \subseteq P \, \text{symmetric} \}.$$

Choose $\bR=(V,r_1,\ldots,r_k) \in \calH'$ such that $\chi(\mathbf{R}) > |\mathcal S|$. Since $\vcspo(\Gamma_\bR \cup \Delta_\bR)$ is tractable and the language is conservative, Theorem \ref{vcspconsthm} gives us a symmetric subset $M_\bR \subseteq P_\bR$, where $P_\bR = \{ (a,b) \in D_\bR^2, a \neq b\}$,  an STP $\sigma = \frac12\chi_\sqcap + \frac12 \chi_\sqcup$ on $M_\bR$, and an MJN $\mu = \frac13 \chi_{F_1} + \frac13 \chi_{F_2} + \frac13 \chi_{F_3}$ on $P_\bR \setminus M_\bR$, such that both $\sigma$ and $\mu$ are fractional polymorphisms of $\Gamma_\bR \cup \Delta_\bR$.

For each $v \in V$, we define a symmetric $M_v \subseteq P$ as $$M_v = \{ (a,b) \in P \colon (d_v(a), d_v(b)) \in M_\bR \}.$$ 
(Again, we use the notation from Sec.~\ref{sec:construction'}.) Further, we set
$\sigma_v = \frac12\chi_{\sqcap_v} + \frac12 \chi_{\sqcup_v}$ where $\sqcap_v \colon D^2 \to D$ is given by 
$a \sqcap b = d( d_v(a) \sqcap d_v(b))$ and $\sqcup_v \colon D^2 \to D$ is defined analogously. 
And finally $\mu_v = \frac13 \chi_{{F_1}_v} + \frac13 \chi_{{F_2}_v} + \frac13 \chi_{{F_3}_v}$, where for $i = 1,2,3$ we define ${F_i}_v \colon D^3 \to D$ as $${F_i}_v(a,b,c) = d(F_i( d_v(a), d_v(b), d_v(c))).$$

It is easily seen that $\sigma_v$ is an STP on $M_v$ and $\mu_v$ is an MJN on $P \setminus M_v$ and thus $(\sigma_v, \mu_v, M_v) \in \calS$.

We use the triples $(\sigma_v, \mu_v, M_v)$ as colors for elements of $V$. Since $\chi(\mathbf{R}) > |\mathcal{S}|$, 
this coloring is improper and therefore there exists a triple $(\tilde\sigma,\tilde\mu,\tilde M) \in \mathcal{S}$ and tuples
$\mathbf{v}^i = (v^i_1, \ldots, v^i_{\ar(r_i)}) \in r_i$  for each $i\in[k]$ such that $(\sigma_v, \mu_v, M_v) = (\tilde\sigma,\tilde\mu,\tilde M)$ for all $v \in \{v^i_j \colon i \in [k], j \in [\ar(r_i)]  \}$.

Next, we show that $\tilde\sigma$ and $\tilde\mu$ are fractional polymorphisms of $\Gamma$. This finishes the proof, since then $\vcspo(\Gamma)$ is tractable by Theorem \ref{vcspconsthm}.

Let us show that $\tilde\sigma = \frac12 \chi_{\tilde\sqcap} + \frac12 \chi_{\tilde\sqcup}$ is admitted by a cost function $f_i \in \Gamma$ for $i\in[k]$. Take $\mathbf{a} = (a_1, \dots, a_p) , \mathbf{b} = (b_1, \dots, b_p) \in \dom f_i$ where $p$ is the arity of $f_i$. First note that
$$
\mathbf{a}\;\tilde\sqcap\; \mathbf{b}
= (a_1 \sqcap_{v^i_1} b_1, \ldots,  a_p \sqcap_{v^i_p} b_p) 
= ( d( d_{v^i_1}(a_1) \sqcap d_{v^i_1}(b_1) ), \ldots,  d( d_{v^i_p}(a_p) \sqcap d_{v^i_p}(b_p))).
$$
Therefore we get
$$
f_i(\mathbf{a}\;\tilde\sqcap\; \mathbf{b}) 
= f_i( d( d_{v^i_1}(a_1) \sqcap d_{v^i_1}(b_1) ), \ldots,  d( d_{v^i_p}(a_p) \sqcap d_{v^i_p}(b_p) )) 
= f_i^{\mathbf{v}^i}( d_{\mathbf{v}^i}(\mathbf{a}) \sqcap d_{\mathbf{v}^i}(\mathbf{b})  )
$$
where in the second equality we used Lemma \ref{vcsptriv_lemma}. Since we have similar equalities for $\tilde\sqcup$ and since $f_i^{\mathbf{v}^i} \in \Gamma_\bR$ admits $\sigma$, we get the sought
\begin{align*}
f_i(\mathbf{a}\;\tilde\sqcap\; \mathbf{b})+f_i(\mathbf{a}\;\tilde\sqcup\; \mathbf{b})
&= f_i^{\mathbf{v}^i}( d_{\mathbf{v}^i}(\mathbf{a}) \sqcap d_{\mathbf{v}^i}(\mathbf{b})  )
+f_i^{\mathbf{v}^i}( d_{\mathbf{v}^i}(\mathbf{a}) \sqcup d_{\mathbf{v}^i}(\mathbf{b})  ) \\ 
&\leq f_i^{\mathbf{v}^i}(d_{\mathbf{v}^i}(\mathbf{a})) + f_i^{\mathbf{v}^i}(d_{\mathbf{v}^i}(\mathbf{b})) = f_i(\mathbf{a}) + f_i(\mathbf{b})
\end{align*}
where in the last equality we used Lemma \ref{vcsptriv_lemma}. Hence $\tilde\sigma$ is admitted by $\Gamma$ and for analogous reasons also $\tilde\mu$ is admitted by $\Gamma$.

\end{proof}

\section*{Acknowledgements}
We thank Andrei Krokhin for helpful comments on the manuscript.
This work was supported by the European Research Council under the European Unions Seventh Framework Programme (FP7/2007-2013)/ERC grant agreement no 616160.

\bibliographystyle{plain}
\bibliography{lit}

\newcommand{\noopsort}[1]{}
\begin{thebibliography}{10}

\bibitem{appel1977}
K.~Appel and W.~Haken.
\newblock Every planar map is four colorable. {P}art i: Discharging.
\newblock {\em Illinois J. Math.}, 21(3):429--490, 09 1977.

\bibitem{BartoKozikLics10}
L.~Barto and M.~Kozik.
\newblock New conditions for {T}aylor varieties and {CSP}.
\newblock In {\em Proceedings of the 25th Annual {IEEE} Symposium on Logic in
  Computer Science, {LICS} 2010, 11-14 July 2010, Edinburgh, United Kingdom},
  pages 100--109, 2010.

\bibitem{barto09:siam}
L.~Barto, M.~Kozik, and T.~Niven.
\newblock The {C}{S}{P} dichotomy holds for digraphs with no sources and no
  sinks (a positive answer to a conjecture of {B}ang-{J}ensen and {H}ell).
\newblock {\em {SIAM} Journal on Computing}, 38(5):1782--1802, 2009.

\bibitem{bulatov06:3-elementjacm}
A.~Bulatov.
\newblock A dichotomy theorem for constraint satisfaction problems on a
  3-element set.
\newblock {\em Journal of the ACM}, 53(1):66--120, 2006.

\bibitem{bulatov11:conservative}
A.~Bulatov.
\newblock Complexity of conservative constraint satisfaction problems.
\newblock {\em ACM Transactions on Computational Logic}, 12(4), 2011.
\newblock Article 24.

\bibitem{Bulatov03:multi-sorted}
A.~Bulatov and P.~Jeavons.
\newblock An algebraic approach to multi-sorted constraints.
\newblock In {\em {CP'03}}, volume 2833 of {\em LNCS}, pages 183--198, 2003.

\bibitem{bulatov05:classifying}
A.~Bulatov, A.~Krokhin, and A.~Jeavons.
\newblock Classifying the {C}omplexity of {C}onstraints using {F}inite
  {A}lgebras.
\newblock {\em {SIAM} Journal on Computing}, 34(3):720--742, 2005.

\bibitem{DBLP:conf/cp/BulinDJN13}
J.~Bul\'in, D.~Delic, M.~Jackson, and T.~Niven.
\newblock On the reduction of the {CSP} dichotomy conjecture to digraphs.
\newblock In Christian Schulte, editor, {\em CP}, volume 8124 of {\em Lecture
  Notes in Computer Science}, pages 184--199. Springer, 2013.

\bibitem{Cohen06:algebraic}
D.~Cohen, M.~Cooper, and P.~Jeavons.
\newblock An algebraic characterisation of complexity for valued constraints.
\newblock In {\em {CP'06}}, volume 4204 of {\em LNCS}, pages 107--121, 2006.

\bibitem{cohen06:complexitysoft}
D.~Cohen, M.~C. Cooper, P.~Jeavons, and A~Krokhin.
\newblock The {C}omplexity of {S}oft {C}onstraint {S}atisfaction.
\newblock {\em Artificial Intelligence}, 170(11):983--1016, 2006.

\bibitem{cccjz13:sicomp}
David~A. Cohen, Martin~C. Cooper, P\'aid\'i Creed, Peter Jeavons, and Stanislav
  {\noopsort{ZZ}\v{Z}}ivn\'y.
\newblock An algebraic theory of complexity for discrete optimisation.
\newblock {\em SIAM Journal on Computing}, 2013.

\bibitem{Cook:1971}
Stephen~A. Cook.
\newblock The complexity of theorem-proving procedures.
\newblock In {\em Proceedings of the Third Annual ACM Symposium on Theory of
  Computing}, STOC '71, pages 151--158, New York, NY, USA, 1971. ACM.

\bibitem{cz11:ai}
Martin~C. Cooper and Stanislav {\noopsort{ZZ}\v{Z}}ivn\'y.
\newblock Hybrid tractability of valued constraint problems.
\newblock {\em Artificial Intelligence}, 175(9-10):1555--1569, 2011.

\bibitem{Erdös1966149}
P.~Erd\H{o}s.
\newblock On the construction of certain graphs.
\newblock {\em Journal of Combinatorial Theory}, 1(1):149 -- 153, 1966.

\bibitem{feder98:monotone}
Tom\'as Feder and Moshe~Y. Vardi.
\newblock The {C}omputational {S}tructure of {M}onotone {M}onadic {S{N}{P}} and
  {C}onstraint {S}atisfaction: {A} {S}tudy through {D}atalog and {G}roup
  {T}heory.
\newblock {\em {SIAM} Journal on Computing}, 28(1):57--104, 1998.

\bibitem{geiger1968}
David Geiger.
\newblock Closed systems of functions and predicates.
\newblock {\em Pacific J. Math.}, 27(1):95--100, 1968.

\bibitem{GreenCohen:AI08}
Martin~J. Green and David~A. Cohen.
\newblock Domain permutation reduction for constraint satisfaction problems.
\newblock {\em Artificial Intelligence}, 172(8-9):1094--1118, 2008.

\bibitem{Grohe:2007}
Martin Grohe.
\newblock The complexity of homomorphism and constraint satisfaction problems
  seen from the other side.
\newblock {\em J. ACM}, 54(1):1:1--1:24, March 2007.

\bibitem{Grotschel88:geometric}
M.~Gr\"{o}tschel, L.~Lov\'asz, and A.~Schrijver.
\newblock {\em Geometric Algorithms and Combinatorial Optimization}.
\newblock Springer-Verlag, New York, 1988.

\bibitem{Hell}
Pavol Hell and Jaroslav Ne\v{s}et\v{r}il.
\newblock On the complexity of h-coloring.
\newblock {\em Journal of Combinatorial Theory, Series B}, 48(1):92 -- 110,
  1990.

\bibitem{jeavons14:beatcs}
P.~Jeavons, A.~Krokhin, and S.~\v{Z}ivn\'y.
\newblock The complexity of valued constraint satisfaction.
\newblock {\em Bulletin of the {EATCS}}, 113:21--55, 2014.

\bibitem{Jeavons:1998}
Peter Jeavons.
\newblock On the algebraic structure of combinatorial problems.
\newblock {\em Theor. Comput. Sci.}, 200(1-2):185--204, June 1998.

\bibitem{Jegou93a}
Philippe J\'egou.
\newblock Decomposition of domains based on the micro-structure of finite
  constraint-satisfaction problems.
\newblock In Richard Fikes and Wendy~G. Lehnert, editors, {\em AAAI}, pages
  731--736. AAAI Press / The MIT Press, 1993.

\bibitem{kz13:jacm}
V.~Kolmogorov and S.~{\noopsort{ZZ}\v{Z}}ivn\'y.
\newblock The complexity of conservative valued {C}{S}{P}s.
\newblock {\em Journal of the ACM}, 60(2), 2013.
\newblock Article 10.

\bibitem{kuznetsov}
A.~V. Kuznetsov.
\newblock Algebra of logic and their generalizations.
\newblock In S.~Janovskaya, editor, {\em Mathematics in USSR for 40 years},
  volume~1, pages 105--115. Fizmatgiz Moscow, 1959.

\bibitem{mader67}
W.~Mader.
\newblock Homomorphieeigenschaften und mittlere kantendichte von graphen.
\newblock {\em Mathematische Annalen}, 174(4):265--268, 1967.

\bibitem{MarotiMcKenzie}
M.~Mar\'oti and R.~McKenzie.
\newblock Existence theorems for weakly symmetric operations.
\newblock {\em Algebra universalis}, 59(3--4):463--489, October 2008.

\bibitem{nesetril:minorclosed}
J.~Ne\v{s}et\v{r}il and P.~Ossona~de Mendez.
\newblock Colorings and homomorphisms of minor closed classes.
\newblock In Boris Aronov, Saugata Basu, J\'anos Pach, and Micha Sharir,
  editors, {\em Discrete and Computational Geometry}, volume~25 of {\em
  Algorithms and Combinatorics}, pages 651--664. Springer Berlin Heidelberg,
  2003.

\bibitem{Post41}
Emil~L. Post.
\newblock {\em {On The Two-Valued Iterative Systems of Mathematical Logic}}.
\newblock Princeton University Press, 1941.

\bibitem{schaefer78:complexity}
T.~J. Schaefer.
\newblock The {C}omplexity of {S}atisfiability {P}roblems.
\newblock In {\em Proceedings of the 10th {A}nnual {A}{C}{M} {S}ymposium on
  {T}heory of {C}omputing ({S}{T}{O}{C}'78)}, pages 216--226. ACM, 1978.

\bibitem{Siggers}
M.~H. Siggers.
\newblock A strong {M}al'cev condition for locally finite varieties omitting
  the unary type.
\newblock {\em Algebra universalis}, 64(1--2):15--20, October 2010.

\bibitem{swarts}
Jacobus~Stephanus Swarts.
\newblock {\em The complexity of digraph homomorphisms: Local tournaments,
  injective homomorphisms and polymorphisms}.
\newblock PhD thesis, University of Victoria, Canada, 2008.

\bibitem{Takhanov10adichotomy}
Rustem~S. Takhanov.
\newblock A dichotomy theorem for the general minimum cost homomorphism
  problem.
\newblock In {\em In Proceedings of the 27th International Symposium on
  Theoretical Aspects of Computer Science (STACS)}, pages 657--668, 2010.

\bibitem{Thapper15:Sherali}
J.~Thapper and S.~\v{Z}ivn\'y.
\newblock Sherali-{A}dams relaxations for valued {CSP}s.
\newblock Technical report, arXiv:1502.05301, 2015.

\bibitem{tz13:stoc}
Johan Thapper and Stanislav {\noopsort{ZZ}\v{Z}}ivn\'y.
\newblock The complexity of finite-valued {C}{S}{P}s.
\newblock In {\em Proceedings of the 45th ACM Symposium on the Theory of
  Computing (STOC'13)}, pages 695--704. ACM, 2013.

\end{thebibliography}

\end{document}